\documentclass[prd,onecolumn,preprintnumbers,nofootinbib]{revtex4}
\usepackage[plainpages=false, colorlinks=true, anchorcolor=blue,
linkcolor=blue, citecolor=blue, bookmarks=false]{hyperref}

\usepackage{graphicx}
\usepackage{dcolumn}
\usepackage{bm}
\usepackage{amsmath, amssymb, amsthm}
\usepackage{float}
\usepackage{enumitem}
\usepackage{booktabs}
\usepackage{caption}
\newcommand{\rthis}[1]{\textcolor{black}{#1}}
\usepackage{subcaption}

\begin{document}
\raggedbottom


\title{\textbf{A comprehensive assessment of weak-lensing inferred circular velocity profiles of isolated galaxies}}

\author{Sri Devaki \surname{Meduri}}
\affiliation{Department of Physics, Indian Institute of Technology Hyderabad,
             Kandi, Sangareddy, Telangana 502284, India}
\altaffiliation{E-mail: ep23btech11030@iith.ac.in}

\author{Shantanu Desai}
\altaffiliation{E-mail: shntn05@gmail.com}
\affiliation{Department of Physics, Indian Institute of Technology Hyderabad,
             Kandi, Sangareddy, Telangana 502284, India}

\begin{abstract}
We fit the circular velocity data, derived from weak lensing observations of isolated galaxies selected from the KiDS survey  in four baryonic mass bins,  using three different dark matter profiles. 
These include NFW, Burkert, and pseudo-isothermal profile. We find that the NFW and Burkert profiles cannot adequately fit the circular velocity data for all the four baryonic  mass bins, whereas only the pseudo-isothermal profile can fit the data for all the  four bins. 
\end{abstract}

\maketitle

\section{Introduction}
We have known for nearly a century that the dominant contribution to the mass of galaxies comes from dark matter~\cite{Hooper,Bosma}. A wide range of cosmological observations has led to the concordance $\Lambda$CDM model~\cite{Planck18}, in which  this dark matter constitutes about 25\% of the total matter-energy density of the universe and is non-relativistic at the time of structure formation~\cite{Rees84}, 70\% is made up of dark energy and baryons constitute the remaining 5\%~\cite{Huterer,Planck18}.

However, there have been  persistent problems with the aforementioned $\Lambda$CDM cosmological model. Some of these problems with the standard model include the core-cusp, missing satellites  and too big-to-fail problem~\citep{Weinberg15,Bullock,Morgan}, Hubble constant tension~\citep{Divalentino}, failure to detect cold dark matter candidates in laboratory-based experiments~\citep{Merritt}, the lithium-7 problem in Big-Bang nucleosynthesis~\citep{Fields}, etc.
A review of   some of the problems associated with the $\Lambda$CDM model, as well as possible alternatives can be found elsewhere (~\cite{Periv,Abdalla22,Peebles22,alternatives,Cosmoverse} and references therein).

One such anomaly which has generated considerable attention over the last decade is the radial acceleration relation (RAR), established with high statistical significance  using the Spitzer photometry and accurate rotation curves (SPARC) galaxy sample~\cite{McGaugh16}. The RAR expresses a tight, universal correlation between the observed centripetal acceleration $g_\mathrm{obs}$ and the Newtonian acceleration predicted from the baryonic mass distribution alone, $g_\mathrm{bar}$:

\begin{equation}
    g_\mathrm{obs}
    = \frac{g_\mathrm{bar}}{1 - \exp\!\left(-\sqrt{g_\mathrm{bar}/g_\dagger}\right)},
    \label{eq:rar}
\end{equation}
where $g_\dagger \approx 1.2 \times 10^{-10}\ \mathrm{m\,s^{-2}}$ \cite{MisteleJCAP} is an empirical acceleration scale. The relation provides an excellent description of galaxies spanning nearly five orders of magnitude in stellar mass --- from dwarf irregulars to massive spirals --- with an intrinsic scatter of only $\sim 0.13$\,dex~\cite{2017ApJ...836..152L}. The tight coupling it encodes between baryonic and dark matter components implies that the baryonic mass distribution alone uniquely predicts the total gravitational field, irrespective of whether a given galaxy is baryon- or dark-matter-dominated~\cite{Mcgaugh14}. However, subsequently  it has also been shown that the RAR is not universal across all dark matter dominated systems~\cite{Salucci18,ChanDesai,Gopika21,Pradyumna,Chan24,Bilek26}.

The physical interpretation of the RAR is an active topic of debate. Within the $\Lambda$CDM framework, it is understood as an emergent statistical property of the galaxy formation process, wherein the functional form of the relation reflects the interplay between halo structure and the efficiency of stellar mass assembly or dynamical interactions between baryons and dark matter~\cite{Aseem,Mayer}. In Modified Newtonian Dynamics (MOND) and related frameworks, by contrast, it is regarded as a direct consequence of a modified law of gravitation, with $g_\dagger$ playing the role of a fundamental constant of nature~\cite{Banik21,Famaey25,Desmond25}. 

 A related anomaly is the Baryonic Tully--Fisher Relation (BTFR), which  is an empirical power-law correlation between the total baryonic mass $M_b = M_\star + M_\mathrm{gas}$ and the asymptotic flat rotation velocity $V_\mathrm{flat}$:
\begin{equation}
    M_b = A\,V_\mathrm{flat}^4,
    \label{eq:btfr}
\end{equation}
with $A \approx 50\,M_\odot\,(\mathrm{km\,s^{-1}})^{-4}$ \cite{McGaugh2000,2012LRR....15...10F}. The relation holds over nearly five decades in baryonic mass with remarkably small scatter, encompassing both gas-dominated dwarf galaxies and massive stellar-dominated spirals.
The BTFR is not an independent empirical law but can be  obtained from  the asymptotic limit of RAR evaluated in the regime, where the rotation curve is flat~\cite{Banik21}.  Conversely, one can also derive the RAR slope, scatter, and acceleration scale from the BTFR~\cite{Wheeler19}.

Most recently,~\citet{Mistele} (M24, hereafter) found evidence for flat rotation curves up to 1 Mpc using the KiDS weak lensing data. This result is also consistent with  BTFR,  and is prima facie in tension with the expected dark matter profiles from $\Lambda$CDM simulations, since they are expected to fall off with distance. This result was disputed in \citet{Chan25} (DC25 hereafter), who argued that if a different virial radius as well as stellar to halo mass relation (SHMR) is used, the disagreement between the dark matter profile and flat rotation curves is not stark. In this work, we extend the analysis in DC25 by considering additional SMHM relations and dark matter models.

The manuscript is structured as follows. We discuss \rthis{the analysis methodology along with the results  from M24 and DC25 in Sect.~\ref{sec:methods}}. \rthis{The data used for our analysis is described in Sect.~\ref{sec:data}}. An overview of dark matter profiles used for the analysis can be found in Sect.~\ref{sec:dm}. The different stellar to halo mass relations are reviewed in Sect.~\ref{sec:smhm}, while the different concentration-mass relations used are presented in Sect.~\ref{sec:cm}. We present our results in Sect.~\ref{sec:results} and conclude in Sect.~\ref{sec:conclusions}.

\section{Analysis Methodology}
\label{sec:methods}
M24 considered isolated galaxies from the KiDS survey and measured the circular velocities using the weak  lensing signal. The analysis in M24 was inspired by an earlier work that performed an RAR analysis of KiDS weak lensing data up to 300~kpc~\cite{2021A&A...650A.113B}.

In order to compare the observed circular velocities with expectations from $\Lambda$CDM, M24 performed a fit to the the widely used Navarro-Frenk-White (NFW) profile~\cite{NFW}, dark matter profile assuming a virial mass and a concentration parameter from cosmological N-body simulations.
The dark matter virial mass was obtained  using the stellar mass-halo mass (SMHM) relation from~\cite{Kravtsov18}. Secondly, $M_{200}$ was used as a proxy for virial mass, where $M_{200}$ is the total mass at an overdensity of 200 \rthis{with respect to the critical density for an Einstein De-Sitter universe~\cite{White01}.}  Then, using one specific  mass-concentration relations~\cite{Maccio08}, M24 found that the flat circular velocities are inconsistent with the NFW profile, which predicts a declining  rotation curve. The rotation curves were found to be flat up to 1 Mpc. This result was also shown to be  consistent with BTFR, which was also found to  hold separately for both early and late type galaxies.
M24 also showed that including the  contribution  from the  two-halo term  does not significantly alleviate the discrepancy.

In response to M24 results, DC25 have argued that it is more appropriate to use the SMHM relations from~\citet{Moster13} instead of~\citet{Kravtsov18}. In addition, they have argued that in the Planck best-fit  $\Lambda$CDM model ($\Omega_{\Lambda} \sim 0.7$, $\Omega_{m} \sim 0.3$) the overdensity after virialization is closer to 100~\cite{1998ApJ...495...80B,DelPopolo13}, instead  of 200 as assumed in M24 and most of the literature.  Therefore, DC25 argued that one should use 
$M_{100}$ for the virial mass instead of $M_{200}$ used in M24.
Finally with these two changes, DC25 fitted the  NFW profile~\cite{NFW} to the lensing based circular velocity data  using mass-concentration relations from~\citet{Maccio08}, after rescaling the concentration parameters to an overdensity of 100. With these changes DC25 found that the circular velocity data is consistent with NFW profile at larger radii for some of the stellar mass bins. \rthis{DC25  also  fitted the pseudo-isothermal profile}  using the relations between the NFW scale radius and that of cored profiles.

In this work, we extend the analysis in M24 by considering  additional  mass-concentration  relations along with different SMHM relations when fitting the NFW profile. We also  fit  the pseudo-isothermal and Burkert profiles. Similar to DC25, \rthis{we use $M_{100}$ as a proxy for the virial mass while fitting to the NFW profile.}

\section{Data}
\label{sec:data}
The KiDS survey  is a photometric survey which  covers $1500\ \mathrm{deg}^2$ of sky in four optical bands ($ugri$), with the best atmospheric conditions reserved for deep $r$-band imaging (median $5\sigma$ limiting AB magnitude of 24.9; median seeing $< 0.7$\,arcsec)~\cite{KIDS}. \rthis{The source galaxies were selected from the KiDS-1000 SOM gold catalog~\cite{Hilde21} and lens galaxies from the KiDS bright sample~\cite{Bilicki21}.}
The source and lens galaxy samples were selected using a more stringent isolation criterion, requiring that there be no   neighboring  massive galaxy with more than 10\% of stellar mass within 4 Mpc/$h_{70}$~\cite{MisteleJCAP} compared to the one used  in an earlier work~\cite{2021A&A...650A.113B}, i.e., 3 Mpc/$h_{70}$. Although  the isolation criterion used in~\citet{2021A&A...650A.113B} has been shown to be reliable using $\Lambda$CDM simulations only up to 300 Kpc due to uncertainties in KiDS photometric redshifts, M24 have argued that their more stringent criterion  cut   allows isolated galaxies to be selected up to a much larger distance of  1 Mpc.  However, M24 also point out that 300 kpc is a conservative lower bound for which the isolation criterion can be trusted. More details on the selection of source and lens galaxy samples along with associated systematics  can be found in ~\citet{MisteleJCAP} and M24.

The observed  lensing signal is then  quantified via the tangential shear $\gamma_t$ of background source galaxies around foreground lenses, from which the excess surface density (ESD) profile $\Delta\Sigma(R)$ is constructed as follows:
\begin{equation}
    \Delta\Sigma(R)
    = \Sigma_\mathrm{crit}\,\gamma_t(R)
    = \langle\Sigma\rangle(<R) - \Sigma(R).
\end{equation}
The ESD is subsequently converted to a stacked acceleration profile via the de-projection formula  discussed in M24. The stacked accelerations are then converted to circular velocities using :
\begin{equation}
    V_c(r) = \sqrt{4G\,\Delta\Sigma(r)\,r},
\end{equation}
allowing the reconstruction of rotation curves extending to $\sim 1\ \mathrm{Mpc}$, far beyond the reach of 21~cm or H$\alpha$ kinematic observations. The lens sample was divided into four  mass bins with stellar masses  of $M_{*}=1.25 \times 10^{11}$, $6.66 \times 10^{10}$, $3.46 \times 10^{10}$, $8.47 \times 10^9 M_{\odot}$. \rthis{Using this procedure, we obtain circular velocity profiles for four galaxies extending out to  1 Mpc.}

\section{Dark Matter Profiles used}
\label{sec:dm}
We now discuss the different dark matter profiles used to model the circular velocity rotation curves.
\subsection{NFW Density Profile}

The NFW profile is the universal density structure of collisionless dark matter halos identified in cosmological N-body simulations, which has been found to be independent of underlying cosmology~\cite{NFW}:
\begin{equation}
    \rho(r) = \frac{\rho_s}{(r/r_s)(1 + r/r_s)^2},
    \label{eq:nfw}
\end{equation}
where $\rho_s$ is a characteristic density scale and $r_s$  is the scale radius.  This profile has been widely used to model the dark matter density profiles across a wide range of systems from the Milky Way galaxy to galaxy clusters. Although recent data and simulations have found better agreement  with Einasto or generalized NFW profiles~\cite{Aryan,AseemGopika,Dalui26,Straight} (and references therein), in this work we use the  NFW profile in order to easily compare our results with DC25 and M24.

While fitting to NFW profiles, one usually defines a concentration parameter ($c$) as the ratio of the virial radius ($r_{vir}$) of the halo to its scale radius $c\equiv r_{vir}/r_s$. 
\rthis{According to  the spherical top-hat model, virialization of an object in an Einstein DeSitter universe occurs at an overdensity $\Delta$ (relative to the critical density of $\frac{3H^2}{8\pi G}$) of $18\pi^2 \approx 178$~\cite{Peebles80}. \citet{NFW} rounded off the virialization overdensity to $\Delta=200$, and assumed $r_{vir}=r_{200}$ and $M_{vir}=M_{200}$. $M_{vir}$ was also considered as a proxy for the halo mass in \citet{NFW}\footnote{Note that there are various other definitions of halo mass in the literature which have been reviewed in ~\citet{White01}}. However, for  Planck-based $\Lambda$CDM cosmology, $\Delta \approx 100$~\cite{1998ApJ...495...80B,DelPopolo13}. Therefore, following DC25, we have adopted $M_{100}$ as the virial mass and $r_{100}$ as the virial radius instead of $M_{200}$ and $R_{200}$.}

\citet{NFW} first found an anti-correlation between halo mass and concentration, which  was found to be sensitive to the slope of the matter power spectrum and accretion history. Consequently, numerous mass-concentration ($c-M$, hereafter) relations have been generated through N-body simulations beginning with ~\citet{Bullock01}. The fit to the NFW profile is then usually done by assuming  a parametric $c-M$ relation  from the literature. DC25 used the $c-M$ relation from~\citet{Maccio08}. In this work, we use several other $c-M$ relations, which are discussed  in Sect.~\ref{sec:cm}.

The resulting circular velocity profile \rthis{squared} for NFW profile can be written as follows:~\cite{Chan25}
\begin{equation}
    V^2(R) = \frac{G M_v}{R_v}
    \cdot \frac{1}{y}
    \cdot \frac{\ln(1+cy) - cy/(1+cy)}{\ln(1+c) - c/(1+c)},
    \label{eq:nfw_vc}
\end{equation}
where $y \equiv R/R_v$, $M_v$ is the virial mass, and $R_v$ is the virial radius. We have adopted $M_{100}$ as the virial mass and $R_{100}$ as the virial radius following DC25.  Note that in this expression, the baryonic mass is neglected, since it's contribution negligible compared to the total mass, \rthis{with a maximum ratio of only 3\% compared to the dark matter mass.}

\subsection{Pseudo-isothermal Profile}
The pseudo-isothermal  sphere is an empirical dark matter halo profile characterized by a central constant-density core, in contrast to the cuspy central behavior of the NFW profile~\cite{2007MNRAS.378...41S}. The density profile is given by:
\begin{equation}
    \rho_\mathrm{pISO}(r) = \frac{\rho_0}{1 + \left(r/R_c\right)^2},
    \label{eq:piso_density}
\end{equation}
where $\rho_0$ is the central density and $R_c$ is the core radius. Integrating the density profile over a sphere of radius $R$ and substituting it into $V^2(R) = GM/R$, where $M$ is taken within the radius $R$, yields the circular velocity profile:
\begin{equation}
    V(R) = \sqrt{4\pi G\rho_0 R_c^2
    \left[1 - \frac{R_c}{R}\arctan\!\left(\frac{R}{R_c}\right)\right]},
    \label{eq:piso_vc}
\end{equation}
which rises steeply at small radii and asymptotes to a finite flat velocity $V_\infty = \sqrt{4\pi G\rho_0}R_c$ at large radii, naturally reproducing flat rotation curves. The two free parameters $\rho_0$ and $R_c$ are determined by a direct least-squares fit  to the observed rotation curve data.

\subsection{Burkert Profile}
The Burkert profile \cite{1995ApJ...447L..25B} is an empirical density profile motivated by observed rotation curves of dwarf galaxies, featuring a central core that transitions to a steeper fall-off at large radii:
\begin{equation}
    \rho_\mathrm{B}(r) = \frac{\rho_0\, r_c^3}
    {(r + r_c)(r^2 + r_c^2)},
    \label{eq:burkert_density}
\end{equation}
where $\rho_0$ is the central density and $R_c$ is the core radius. Integrating to obtain the enclosed mass,
\begin{equation}
    M(r) = \pi\rho_0 r_c^3
    \left[\ln\!\left(1 + \frac{r}{r_c}\right)^2
    + \ln\!\left(1 + \frac{r^2}{r_c^2}\right)
    - 2\arctan\!\left(\frac{r}{r_c}\right)\right],
    \label{eq:burkert_mass}
\end{equation}
The circular velocity profile is then given by:
\begin{equation}
    V(R) = \sqrt{\frac{\pi G\rho_0 r_c^3}{R}
    \left[\ln\!\left(1 + \frac{R}{r_c}\right)^2
    + \ln\!\left(1 + \frac{R^2}{r_c^2}\right)
    - 2\arctan\!\left(\frac{R}{r_c}\right)\right]}.
    \label{eq:burkert_vc}
\end{equation}
Unlike the pseudo-isothermal profile, the Burkert circular velocity peaks at a finite radius and declines at large $R$, making it less suited to galaxies with extended flat rotation curves. As in the pseudo-isothermal profile, the two free parameters $\rho_0$ and $r_c$ are determined by direct least-squares fitting. The Burkert profile also provides a good fit to profiles of self-interacting dark matter  and cold dark matter coupled with feedback~\cite{Straight,Dalui26}.

Note that one difference between NFW \rthis{(after assuming a $c-M$ and SMHM relation)}  and pseudo-isothermal/Burkert profile fitting is that the fits to the latter two models do not depend on SMHM relation.  We also note that unlike DC25, we do not use the relation between  $r_c$ and  $r_s$ from ~\cite{Boyarsky09}, while fitting the pseudo-isothermal profile.

\section{Stellar Mass--Halo Mass Relations}
\label{sec:smhm}

In this section we discuss the different SMHM relations used for our analysis.

\subsection{Moster SMHM Relation}
The Moster relation \cite{Moster13} maps the stellar mass to halo mass through a double power-law parametrization calibrated by abundance matching:
\begin{equation}
    \frac{M_\star}{M_\mathrm{halo}}
    = \frac{2N}{\left(M_\mathrm{halo}/M_1\right)^{-\beta}
               + \left(M_\mathrm{halo}/M_1\right)^{\gamma}},
    \label{eq:moster}
\end{equation}
with $M_1 = 10^{11.59}\,M_\odot$, $N = 0.0351$, $\beta = 1.376$, and $\gamma = 0.698$. $M_{\text{halo}}$ is assumed to be $M_{200}$. The double power-law captures the characteristic inefficiency of galaxy formation at both extremes of the halo mass function: both dwarf halos and cluster-scale halos convert a substantially smaller fraction of available baryons into stars relative to halos near the characteristic mass $M_1 \sim M_{\rm MW}$. The relation was calibrated by matching the observed stellar mass functions from the NYU Value-Added Galaxy Catalog~\cite{2005AJ....129.2562B} and the COSMOS survey~\cite{2007ApJS..172....1S} over the redshift range $0 \leq z \leq 4$~\cite{Moster13}.

The Moster relation has been extensively validated across independent probes. In the Millennium Simulation, it reproduces the observed galaxy two-point correlation functions and clustering statistics~\cite{Moster13}. Weak gravitational lensing analyses confirm consistency between its predicted halo masses and lensing-inferred masses at fixed stellar mass, and it successfully reproduces the velocity dispersion--stellar mass relation for satellite galaxies~\cite{2012ApJ...746...95L,2011MNRAS.410..210M}. 

\subsection{Behroozi SMHM Relation}

The Behroozi relation~\cite{2013ApJ...770...57B} provides an alternative parametrization derived from hydrodynamical simulations and can be described as follows~\cite{2011ApJ...740..102K}:
\begin{equation}
    \log_{10} M_\star = \log_{10}(\epsilon\,M_1) + f(x) - f(0),
    \label{eq:Behroozi}
\end{equation}
where $x = \log_{10}(M_\mathrm{halo}/M_1)$ and
\begin{equation}
    f(x) = -\log_{10}(10^{\alpha x}+1)
    + \delta\,\frac{\left[\log_{10}(1+e^x)\right]^{\gamma}}{1+e^{-x}},
\end{equation}
with $M_1 = 10^{11.514}\,M_\odot$, $\epsilon = 10^{-1.777}$, $\alpha = -1.412$, $\delta = 3.508$, and $\gamma = 0.316$. $M_{\text{halo}}$ is taken to be $M_{200}$.

\rthis{The Behroozi SMHM was calibrated using the Bolshoi simulations~\cite{2011ApJ...740..102K}}, incorporating explicit prescriptions for stellar feedback, supernova-driven outflows, and photo-ionization. It therefore accounts for baryonic modifications to the halo structure and predicts systematically lower stellar masses for a given halo mass relative to pure abundance matching, particularly in the dwarf and intermediate-mass regimes.

This relation has been validated using stellar-to-halo mass ratios inferred from satellite kinematics and weak lensing~\cite{2013ApJ...770...57B}. It  has also been widely  used  in studies of dark matter core formation, the cusp--core transformation in dwarf galaxies, and the `too-big-to-fail' problem. However, the simulation-based calibration ties the relation to the specific feedback prescriptions of the MaGICC suite, which may not be universal. In the present work, the Behroozi relation consistently yields inferior rotation curve fits relative to the Moster relation across all mass scales and $c-M$ combinations, suggesting that its halo mass predictions are systematically less consistent with the observed kinematics of the galaxies in our sample.

\subsection{Cintio SMHM Relation}
The Cintio relation~\cite{2014MNRAS.441.2986D}, derived from the same aforementioned MaGICC simulation suite, extends the hydrodynamical SMHM framework by allowing the characteristic parameters of the double power-law to vary continuously with halo mass.
The equation used for obtaining $M_{\text{halo}}$ is given as follows:
\begin{equation}
    \frac{M_\star}{M_{\text{halo}}} = 2\gamma  \left[ \left(\frac{M_\star}{M_{\text{halo}}}\right)^{-\alpha} + \left(\frac{M_\star}{M_{\text{halo}}}\right)^{\beta}\right]^{-1}
\end{equation}
$\alpha$, $\beta$, $\gamma$ are functions of $X \equiv \log_{10}(M_\star/M_\mathrm{halo})$~\cite{2014MNRAS.441.2986D}. However, the fiducial reference point used is $(\alpha,\beta,\gamma)=(1,3,1)$.
Here also, $M_{\text{halo}}$ is assumed to be $M_{200}$. \rthis{($\alpha,\beta,\gamma)=(1,3,1)$ are the values obtained in~\cite{2014MNRAS.441.2986D} for the NFW profile.}

The key distinction from the Behroozi parametrization is the self-consistent, mass-dependent nature of these parameters. While Behroozi imposes a fixed functional form with globally calibrated constants, the Cintio framework allows the low-mass slope ($\alpha$), the high-mass slope ($\beta$), and the transition sharpness ($\gamma$) to respond to the local feedback environment as a function of the halo mass. This mass-dependent re-parametrization enables the relation to capture the non-self-similar character of stellar mass assembly across the full range from dwarf galaxies to Milky Way-mass halos, where the relative importance of supernova feedback, re-ionization, and halo growth all evolve substantially.

The Cintio relation has been validated against observed stellar mass functions, the Tully--Fisher relation, and the kinematics of Local Group dwarfs \cite{2014MNRAS.441.2986D}, and has been applied in studies of the cusp--core transformation and the `too-big-to-fail' problem~\cite{2011MNRAS.415L..40B}. Because its calibration is anchored to the MaGICC suite \cite{2012MNRAS.424.1275B}, it shares the systematic uncertainties of the Behroozi relation regarding the universality of the adopted feedback prescriptions.

\begin{figure}[H]
\centering
\includegraphics[scale=0.7]{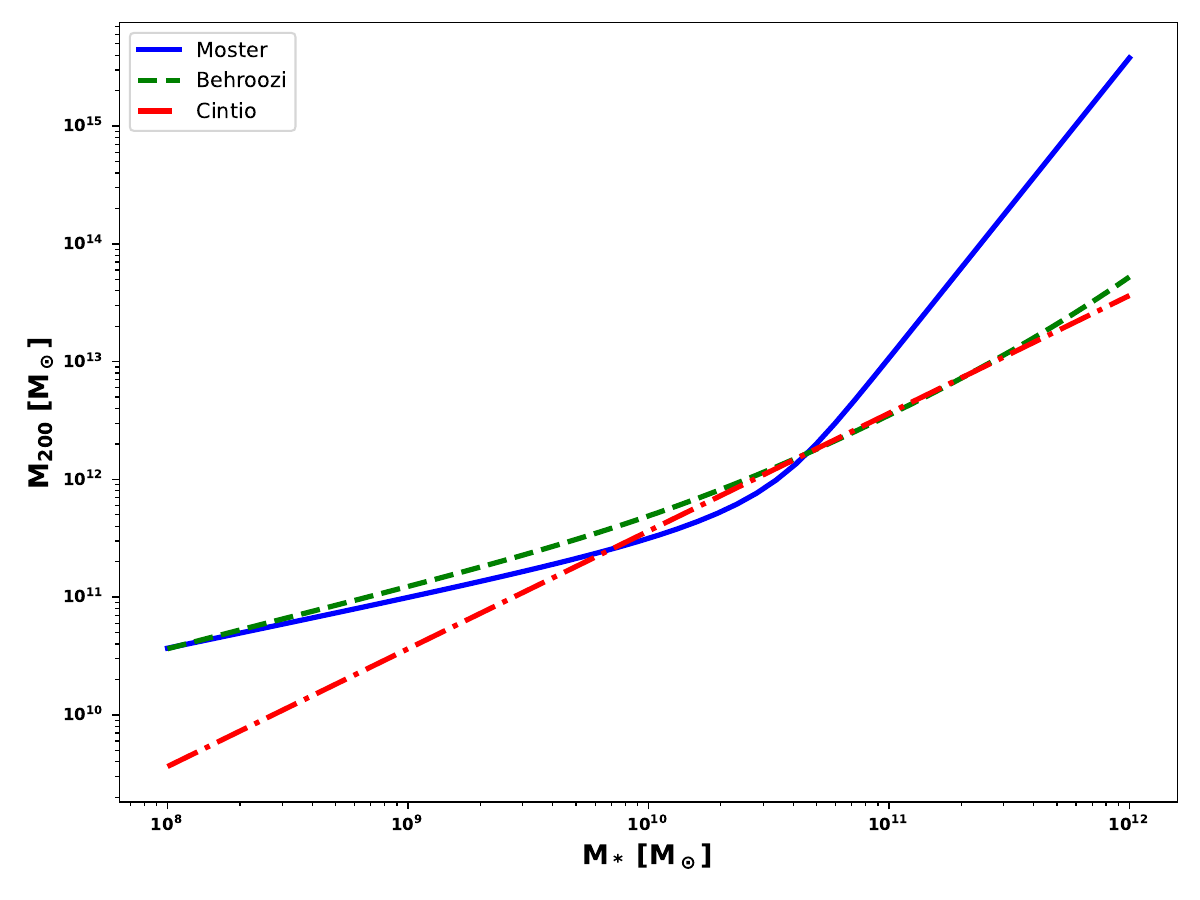}
\caption{Plot of $M_{200}$  vs $M_\star$  for the three SMHM relations used in this analysis while  fitting the NFW profile.}
\label{fig:smhm}
\end{figure}

A comparison of all three SMHM relations described above can be found in Fig.~\ref{fig:smhm}. We find that at low masses the Cintio relation is suppressed compared to other relations, whereas at high masses the Moster relation is elevated compared to other relations. Between $\sim 10^{10}$ and $\sim 10^{11}$, all three SMHM relations are in agreement.

\section{Mass--Concentration Relations}
\label{sec:cm}
Before listing the different $c-M$ relations, needed to fit the NFW profile,  we recap the different halo masses needed for our calculation and how to convert from one to another.

The overdensity mass, $M_\Delta$ is defined as the mass enclosed within the radius at which the mean interior density is equal to $\Delta$ times the critical density $\rho_\mathrm{crit}$.
As discussed earlier, we adopt $M_{100}$ ($\Delta = 100$) as our fiducial definition for the virial mass similar to DC25, which provides a more physically motivated outer boundary than the commonly used $M_{200}$ ($\Delta = 200$).  Since all $c-M$ (and also SMHM) relations  employ $M_{200}$ and $c_{200}$, we need to convert $M_{200}$ to $M_{100}$ and $c_{200}$ to $c_{100}$. 
The conversion from $M_{200}$ to $M_{100}$ exploits the NFW mass function. Defining the mass ratio $M_c \equiv M_{200}/M_\Delta$ and the function
\begin{equation}
    A(c) = \left[\ln(1+c) - \frac{c}{1+c}\right]^{-1}, \qquad c = c_{200},
\end{equation}
The over-density mass satisfies~\cite{Lukic}:
\begin{equation}
    M_c = A(c)
    \left[\ln\!\left(1 + \sqrt[3]{\tfrac{200}{\Delta}}\,M_c\,c\right)
    - \frac{\sqrt[3]{200/\Delta}\,M_c\,c}
           {1+\sqrt[3]{200/\Delta}\,M_c\,c}\right].
    \label{eq:mc}      
\end{equation}

The conversion from $c_{200}$ to $c_{100}$ is done using the following relation~\cite{Coe10}:
\begin{equation}
    c_{100} = 1.298\,c_{200} + 0.246.
    \label{eq:c100}
\end{equation}

For our analyses, six independent $c-M$ relations are used, spanning a range of calibration methodologies. 
We now describe each of these in turn.


\subsection{Duffy et al.}

The Duffy et al.\ \cite{2008MNRAS.390L..64D} concentration--mass relation is calibrated from the Millennium~\cite{2005Natur.435..629S} and Millennium-2~\cite{2011MNRAS.415L..40B} N-body simulations. This relation can be parametrized as:
\begin{equation}
    c_{200} = \alpha\left(\frac{M_{200}}{M_p}\right)^{\beta}(1+z)^{\gamma},
\end{equation}
where $\alpha = 5.71$, $\beta = -0.084$, $\gamma = -0.47$, and $M_p = 2\times10^{12}\,h^{-1}\,M_\odot$ is a pivot mass chosen near the characteristic non-linear mass scale at $z = 0$.

This two-parameter power-law form provides a compact and widely used fitting function that has been validated against both weak-lensing observations~\cite{2007ApJ...667..176G} and satellite kinematics~\cite{2008MNRAS.390L..64D}, making it a standard reference relation in rotation curve and strong-lensing analyses.

\subsection{Diemer \& Joyce}
The Diemer \& Joyce~\cite{2019ApJ...871..168D} relation employs a physics-motivated approach based on the local logarithmic slope $n$ of the linear matter power spectrum:
\begin{equation}
    \frac{c}{\left[g(c)\right]^{(5+n)/6}} = \frac{A(n)}{\nu},
\end{equation}
where $g(c) \equiv \ln(1+c) - c/(1+c)$, $A(n) = \nu_{pe}\,c_{pe}/[g(c_{pe})]^{(5+n)/6}$, $\nu(M,z) = \left(\frac{M}{M_\star(z)}\right)^{\frac{n+3}{6}}$, $M_\star(z)$ is the characteristic mass scale defined at redshift $z$, and $\nu_{pe}$, $c_{pe}$ are pivot-epoch parameters. By encoding the cosmological context through $n$ rather than halo mass alone, this relation captures the dependence of halo structure on the shape of the power spectrum.

\subsection{Prada et al.}
The Prada et al.~\cite{2012MNRAS.423.3018P} relation provides a power-law fit calibrated for galaxy-scale halos, expressing the concentration as a simple function of halo mass and the present-day Hubble parameter:
\begin{equation}
    c_{200} = 7.28\left(\frac{M_{200}\,h}{10^{12}\,M_\odot}\right)^{-0.074}.
\end{equation}
In this form, the mass and redshift dependence of $c(M,z)$ from the underlying cosmology based on how $c$ is defined ~\cite{2012MNRAS.423.3018P}. By calibrating directly against $h$-scaled halo masses, this relation offers a compact, easily applicable description of concentration trends without requiring full evaluation of the peak-height formalism.


\subsection{Diemer \& Kravtsov}
The Diemer \& Kravtsov \cite{2015ApJ...799..108D} relation is calibrated from N-body simulations~\cite{2011ApJ...740..102K} and encodes the dependence of halo concentration on the local logarithmic slope $n_\mathrm{eff}$ of the linear matter power spectrum. The concentration is given by:
\begin{equation}
    c_{200} = \frac{c_\mathrm{min}}{2}
    \left[\left(\frac{\nu}{\nu_0}\right)^{-\alpha}
    + \left(\frac{\nu}{\nu_0}\right)^{\beta}\right],
\end{equation}
where $\nu = \frac{1.686}{\sigma(M,z)}$ is the peak height, $\sigma$ is the rms density fluctuation in a sphere, and the parameters $c_\mathrm{min}$ and $\nu_0$ depend linearly on $n_\mathrm{eff}$. The model reproduces N-body concentrations~\cite{2011ApJ...740..102K} with $\lesssim 5\%$ precision over $0 \leq z \leq 6$.

\subsection{S\'{a}nchez-Conde \& Prada}
The S\'{a}nchez-Conde \& Prada~\cite{2014MNRAS.442.2271S} relation is calibrated from the Bolshoi~\cite{2011ApJ...740..102K} and MultiDark simulations~\cite{2016MNRAS.457.4340K} and is expressed as a degree-5 polynomial in $\ln(M_{200}\,h)$:
\begin{equation}
    c_{200} = \sum_{i=0}^{5} c_i\,\left[\ln(M_{200}\,h)\right]^i,
\end{equation}
with coefficients $(c_0,\ldots,c_5) = (37.5153,\,-1.5093,\,1.636\times10^{-2},\,3.66\times10^{-4},\,-2.892\times10^{-5},\,5.32\times10^{-7})$. Unlike single power-law fits, this high-order polynomial parametrization was introduced to reproduce a feature seen consistently across the Bolshoi and MultiDark $N$-body simulations: the concentration--mass relation does not decrease monotonically with mass but instead flattens and turns upward again at the low-mass end, below $\sim 10^{10}\,h^{-1}\,M_\odot$, before eventually re-steepening toward the smallest resolved halo masses. This non-monotonic behavior, driven by the interplay between halo assembly history and the shape of the matter power spectrum on small scales, cannot be captured by a simple power law, motivating the polynomial form. Because it is calibrated over an extremely wide mass range, spanning from dwarf-galaxy to cluster scales, this $c$--$M$ relation has been extensively used in indirect dark matter searches~\cite{Manna24}, where accurate concentration estimates for low-mass subhalos are essential for predicting annihilation or decay signals.

\subsection{Macci\`{o}}
The Macci\`{o} \cite{Maccio08} relation was obtained using WMAP5 cosmology~\cite{2009ApJS..180..330K}  and is expressed as follows:
\begin{equation}
    \log c_{200} = 0.830 - 0.098\,\log\!\left(\frac{M_{200}}{10^{12}\,h^{-1}M_\odot}\right).
\end{equation}
This relation was derived as a result of the failure of the already existing $c$--$M$ relations in their simulations, which had largely been calibrated on first-year or three-year WMAP cosmologies and systematically failed to reproduce the concentrations measured in halos evolved under the updated WMAP5 parameters, particularly the revised normalization of the matter power spectrum, $\sigma_8$. By recalibrating both the normalization and the logarithmic slope of the $c$--$M$ relation against a new suite of high-resolution $N$-body simulations, Macci\`{o} provided a relation that is internally consistent with WMAP5-era cosmological parameters, underscoring the broader sensitivity of halo concentration to the assumed background cosmology rather than to halo mass alone.

\begin{figure}[h]
\centering
\includegraphics[scale=0.7]{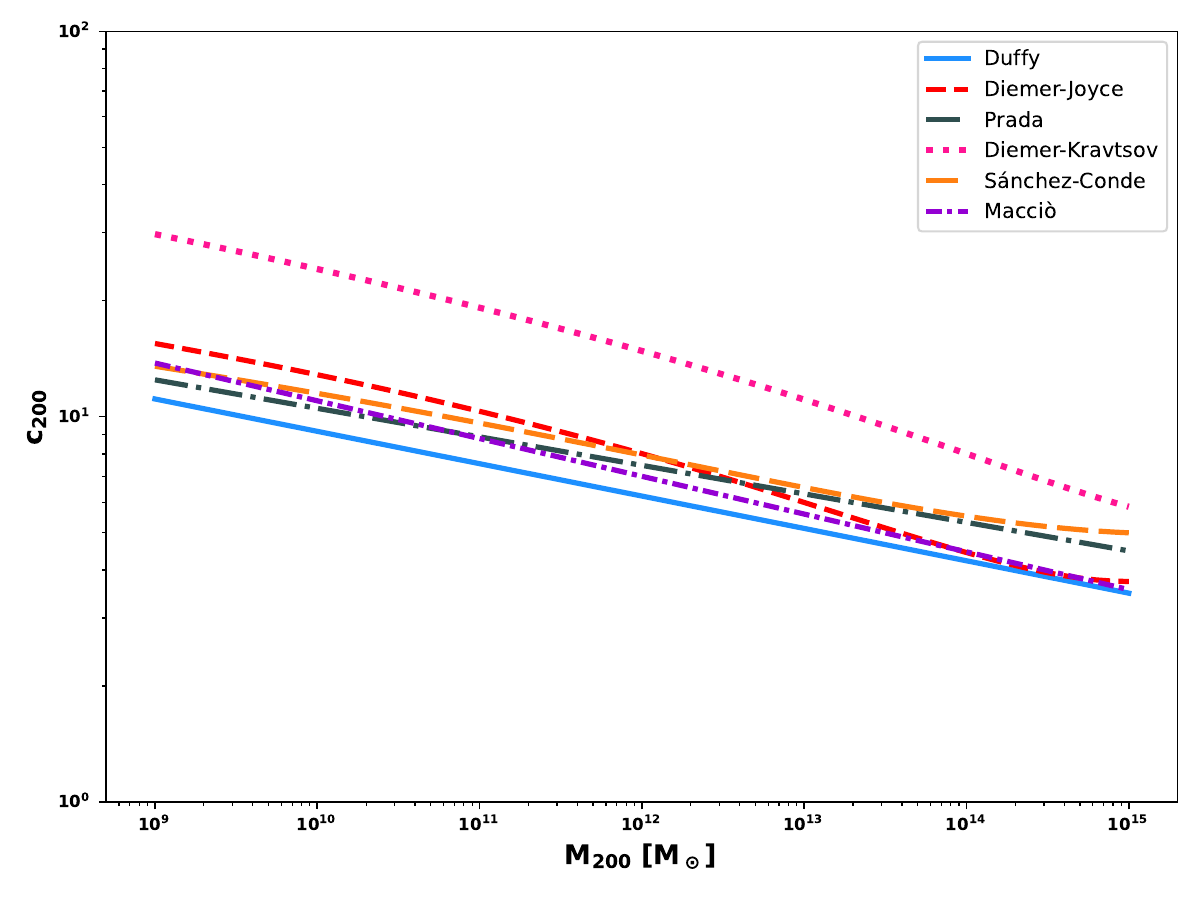}
\caption{Plot of the $c-M$ relations as a function of $M_{200}$($M_\odot$) used for fitting NFW data .}
\label{fig:cM}
\end{figure}

Fig.~\ref{fig:cM}  shows the  plot of $c_{200}$ obtained using all  the aforementioned $c-M$ relations, and how they scale as a function of $M_{200}$. The Kravtsov $c-M$ relation produces significantly high $c_{200}$ while the Duffy relation produces the lowest $c_{200}$ for  the same $M_{200}$.  We find that the Diemer-Kravtsov $c-M$ relation is elevated compared to the others.

\section{Results}
\label{sec:results}
We now present our results. For each dark matter profile (and six $c-M$ relations along with three SMHM relations  for the NFW profile),  we calculate the estimated $V_c(R)$ for each of the four stellar mass bins and overlay them on the  data. 
To quantify the agreement between data and model, we calculate  the reduced $\chi^2$ statistic, defined as follows:
\begin{equation}
    \chi^2_\nu = \frac{1}{N-p}\sum_{i=1}^{N}
    \frac{\left(V_{\mathrm{model},i} - V_{\mathrm{obs},i}\right)^2}{\sigma_i^2},
\end{equation}
where $\sigma_i$ denotes  the observational uncertainties on the circular velocity, $p$ is the number of free parameters in the model and $N$ is the total number of data points. For the NFW profile  $V_{\mathrm{model},i}$ was   calculated using Eq.~\ref{eq:nfw_vc}. For the pseudo-isothermal and  Burkert  profiles, we used  Eq.~\ref{eq:piso_vc} and Eq.~\ref{eq:burkert_vc}, respectively. We note that since we used $c-M$ relation directly for fitting the NFW profile, $p=0$, whereas for the pseudo-isothermal and Burkert profile $p=2$. 

\subsection{NFW profile}

\begin{figure}[p]
\captionsetup[subfigure]{justification=centering, singlelinecheck=false, font=small}
\centering

\begin{subfigure}[t]{0.47\textwidth}
    \includegraphics[width=\linewidth]{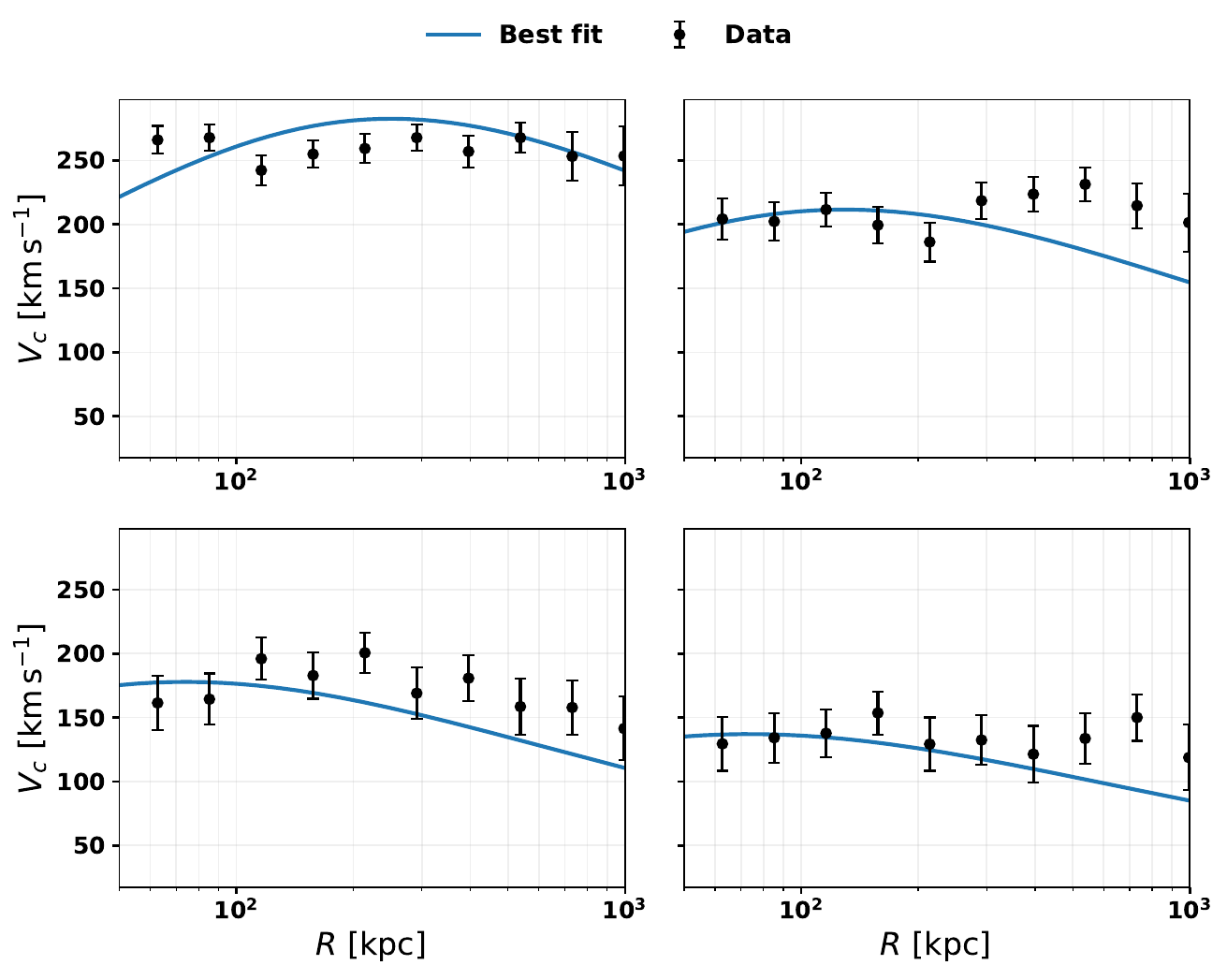}
    \caption{\centering Duffy $c$--$M$\\ 
    $\chi^2_\nu = (2.7,\,3.1,\,1.8,\,1.4)$}
\end{subfigure}
\hfill
\begin{subfigure}[t]{0.47\textwidth}
    \includegraphics[width=\linewidth]{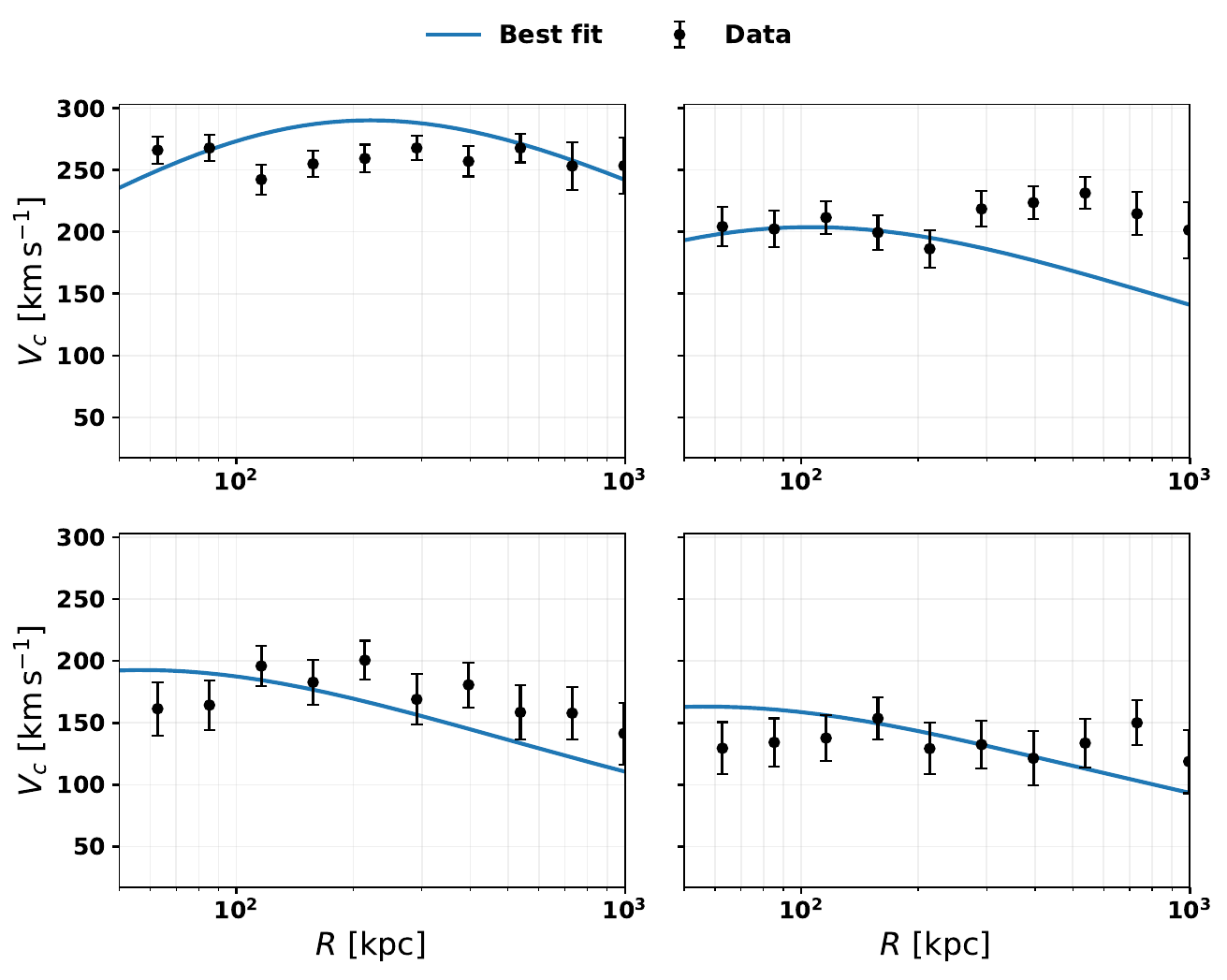}
    \caption{\centering Diemer-Joyce $c$--$M$\\
    $\chi^2_\nu = (3.2,\,5.1,\,2.0,\,1.6)$}
\end{subfigure}

\vspace{0.25cm}

\begin{subfigure}[t]{0.47\textwidth}
    \includegraphics[width=\linewidth]{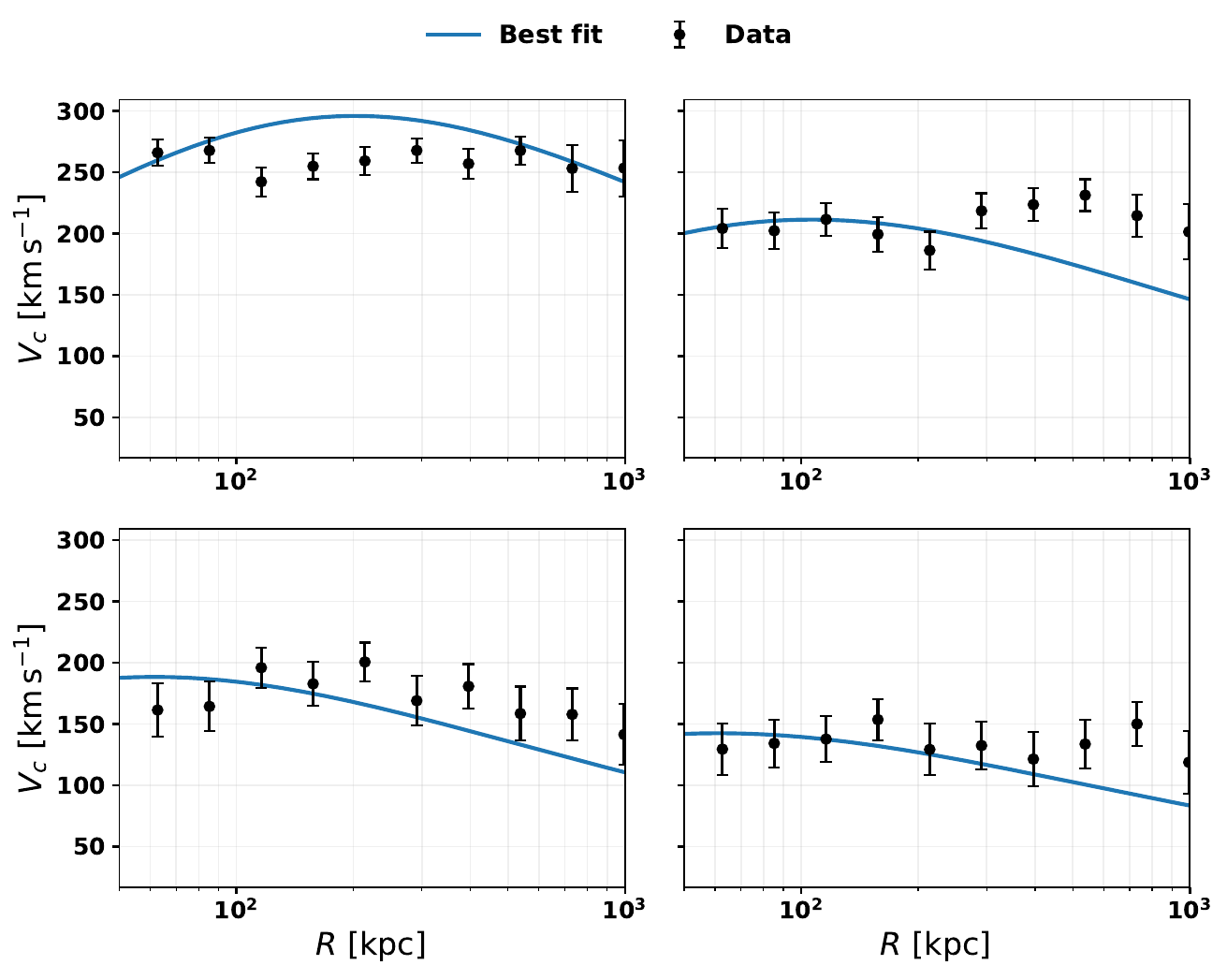}
    \caption{\centering Prada $c$--$M$\\
    $\chi^2_\nu = (4.2,\,4.1,\,1.9,\,1.5)$}
\end{subfigure} 
\hfill
\begin{subfigure}[t]{0.47\textwidth}
    \includegraphics[width=\linewidth]{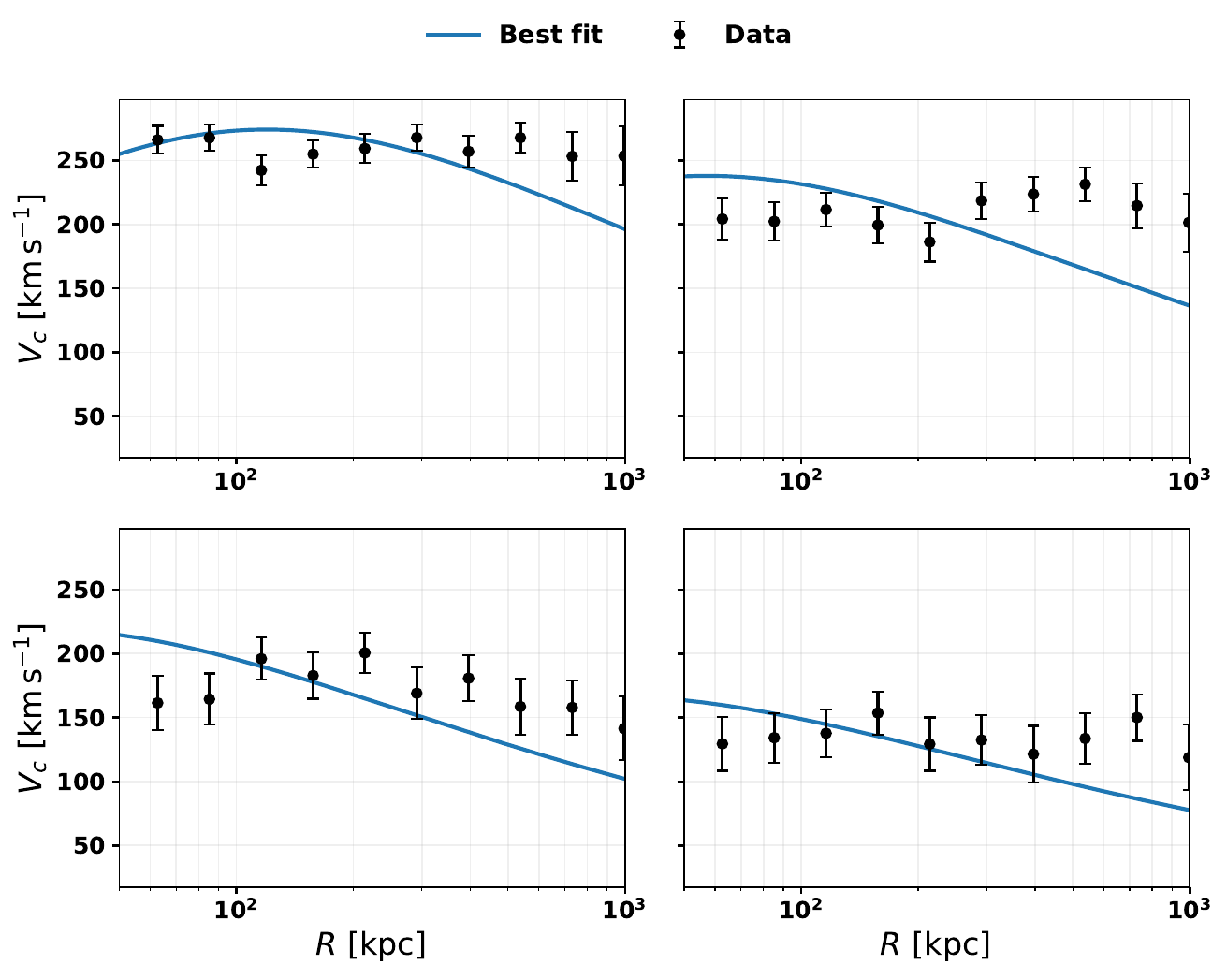}
    \caption{\centering Diemer-Kravtsov $c$--$M$\\
    $\chi^2_\nu = (3.2,\,7.0,\,3.4,\,2.5)$}
\end{subfigure}

\vspace{0.25cm}

\begin{subfigure}[t]{0.47\textwidth}
    \includegraphics[width=\linewidth]{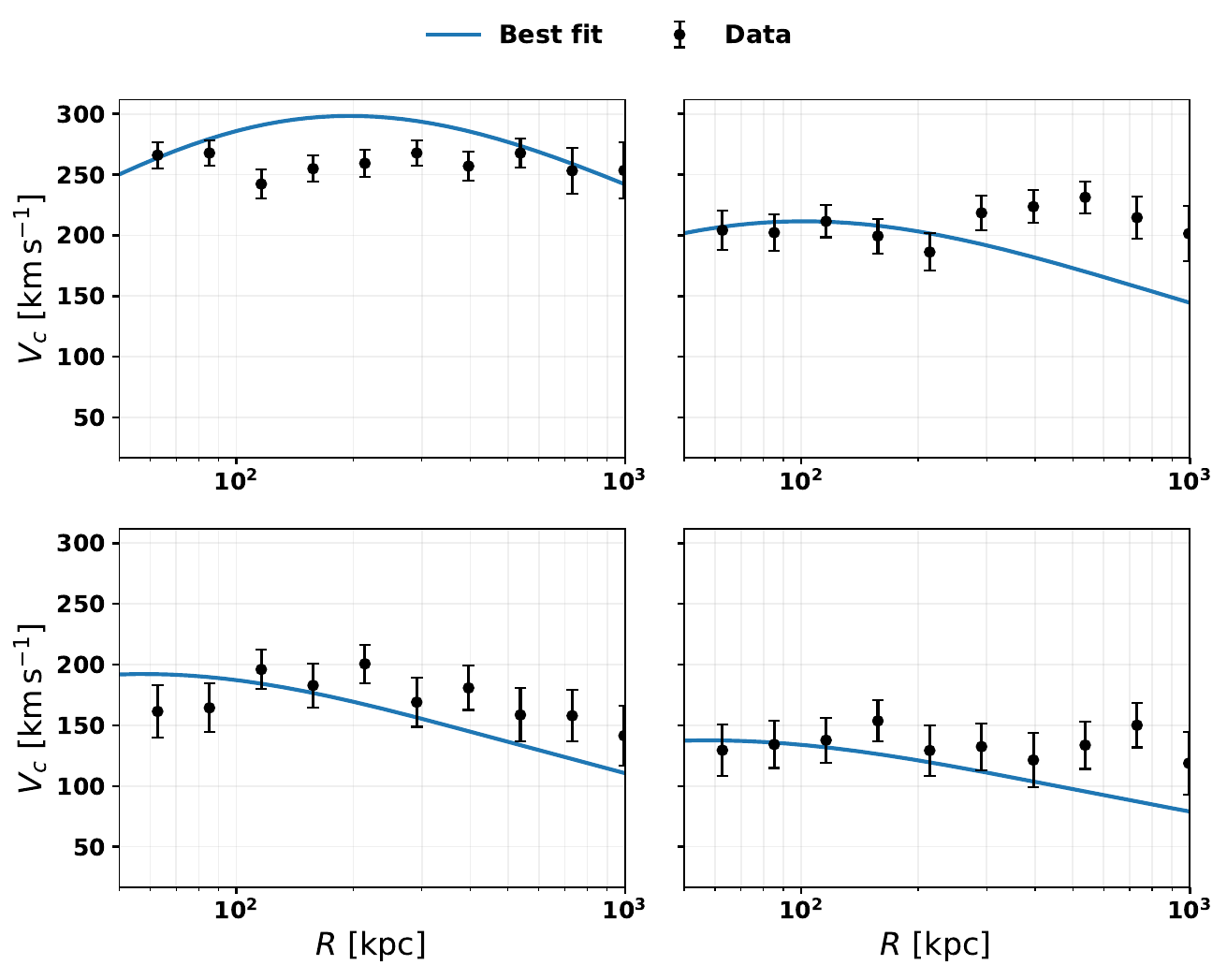}
    \caption{\centering S\'anchez-Conde $c$--$M$\\
    $\chi^2_\nu = (4.8,\,4.4,\,2.0,\,1.9)$}
\end{subfigure}
\hfill
\begin{subfigure}[t]{0.47\textwidth}
    \includegraphics[width=\linewidth]{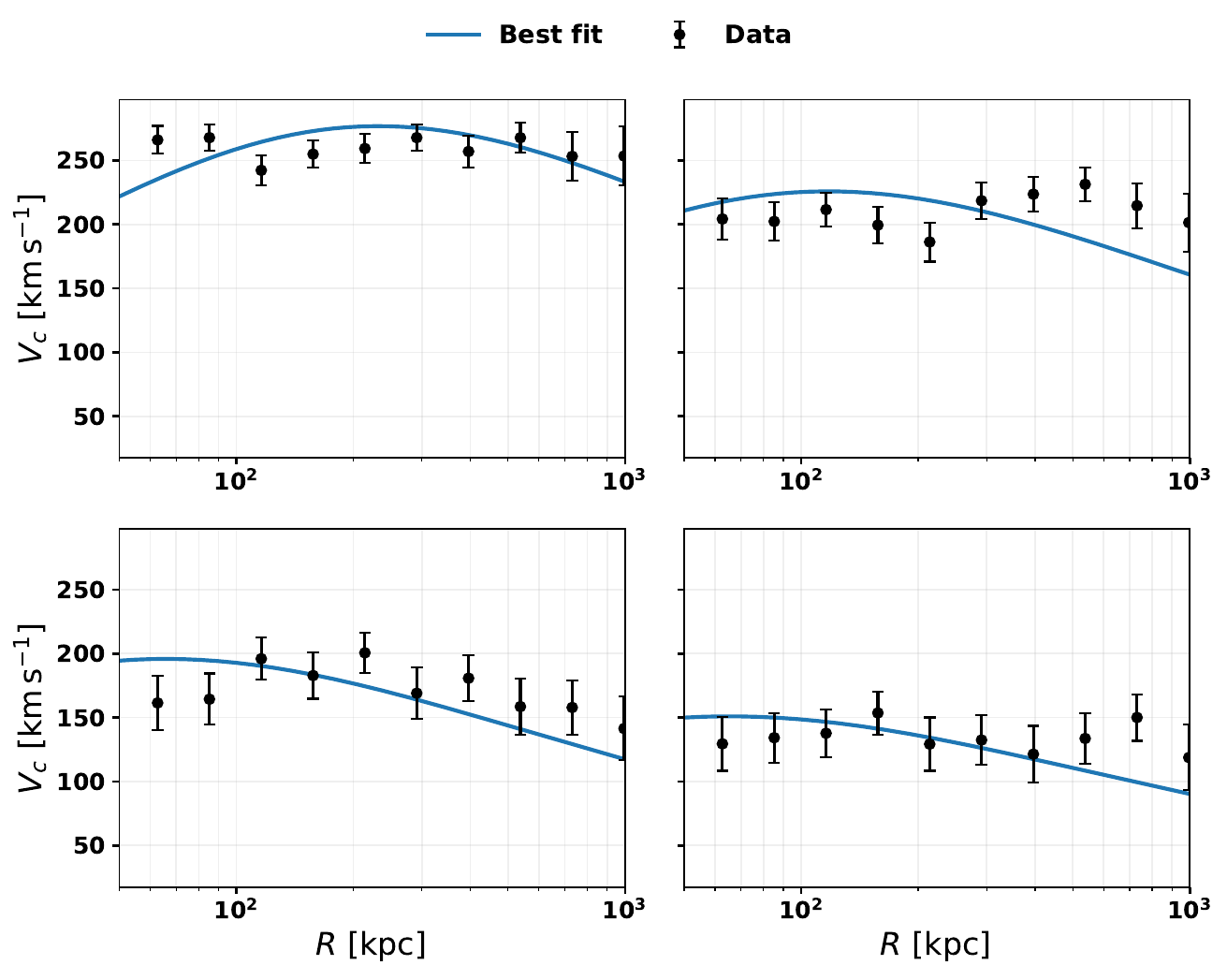}
    \caption{\centering Macci\`{o} $c$--$M$\\
    $\chi^2_\nu = (2.3,\,2.9,\,1.6,\,1.3)$}
\end{subfigure}

\caption{
NFW rotation-curve fits using the Moster SMHM relation combined with six different concentration--mass relations.
Within each panel, the four subplots correspond (clockwise from top left) to stellar masses
$M_\star/M_\odot = 1.25\times10^{11}$,
$6.66\times10^{10}$,
$3.46\times10^{10}$,
and $8.47\times10^{9}$.
The reduced chi-squared values for each mass are listed in the individual subfigure captions.
}
\label{fig:moster_all}
\end{figure}
\begin{figure}[p]
\captionsetup[subfigure]{justification=centering, singlelinecheck=false, font=small}
\centering

\begin{subfigure}[t]{0.47\textwidth}
    \includegraphics[width=\linewidth]{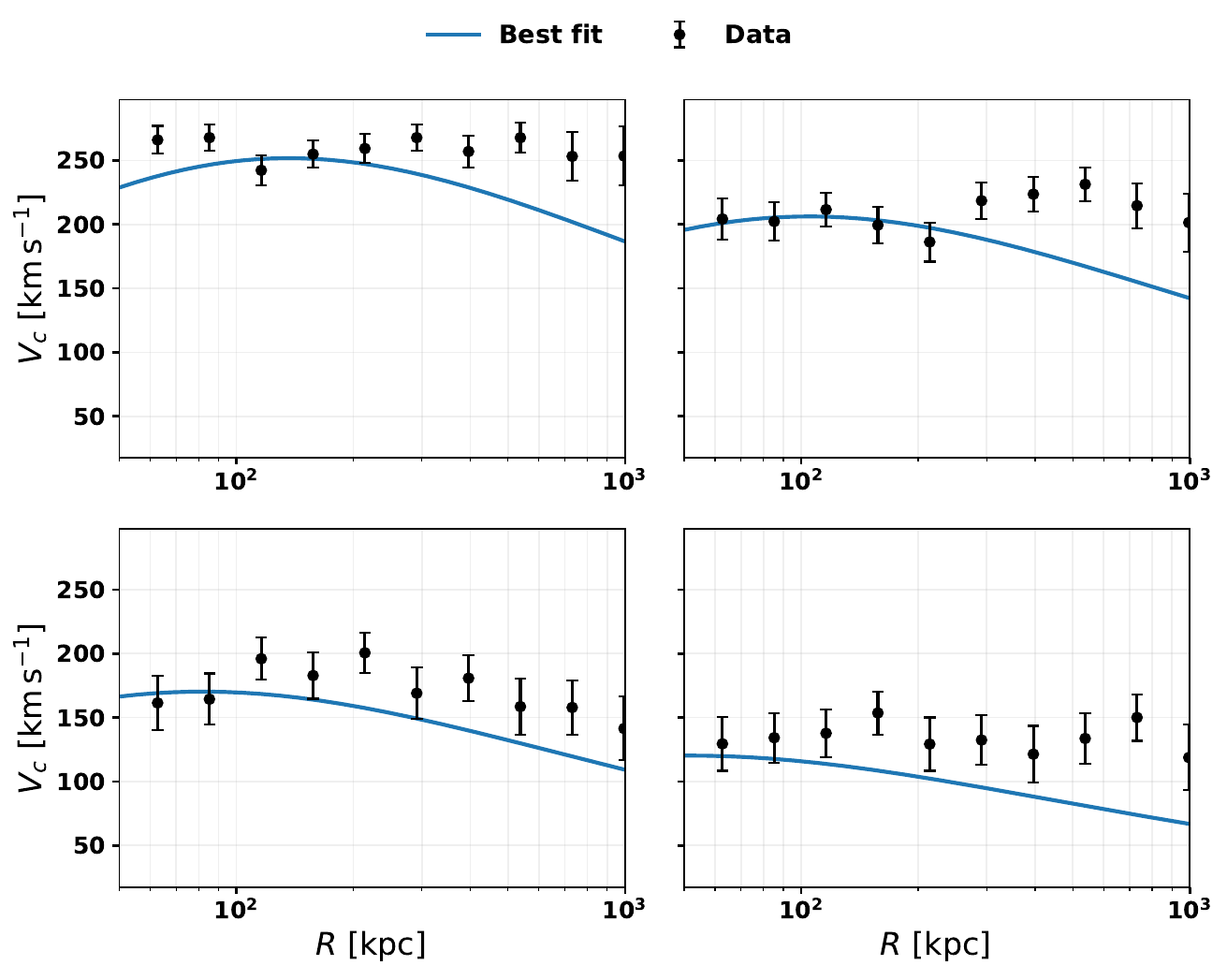}
    \caption{\centering Duffy $c$--$M$: 
    $\chi^2_\nu = (5.1,\,4.8,\,2.0,\,3.6)$}
\end{subfigure}
\hfill
\begin{subfigure}[t]{0.47\textwidth}
    \includegraphics[width=\linewidth]{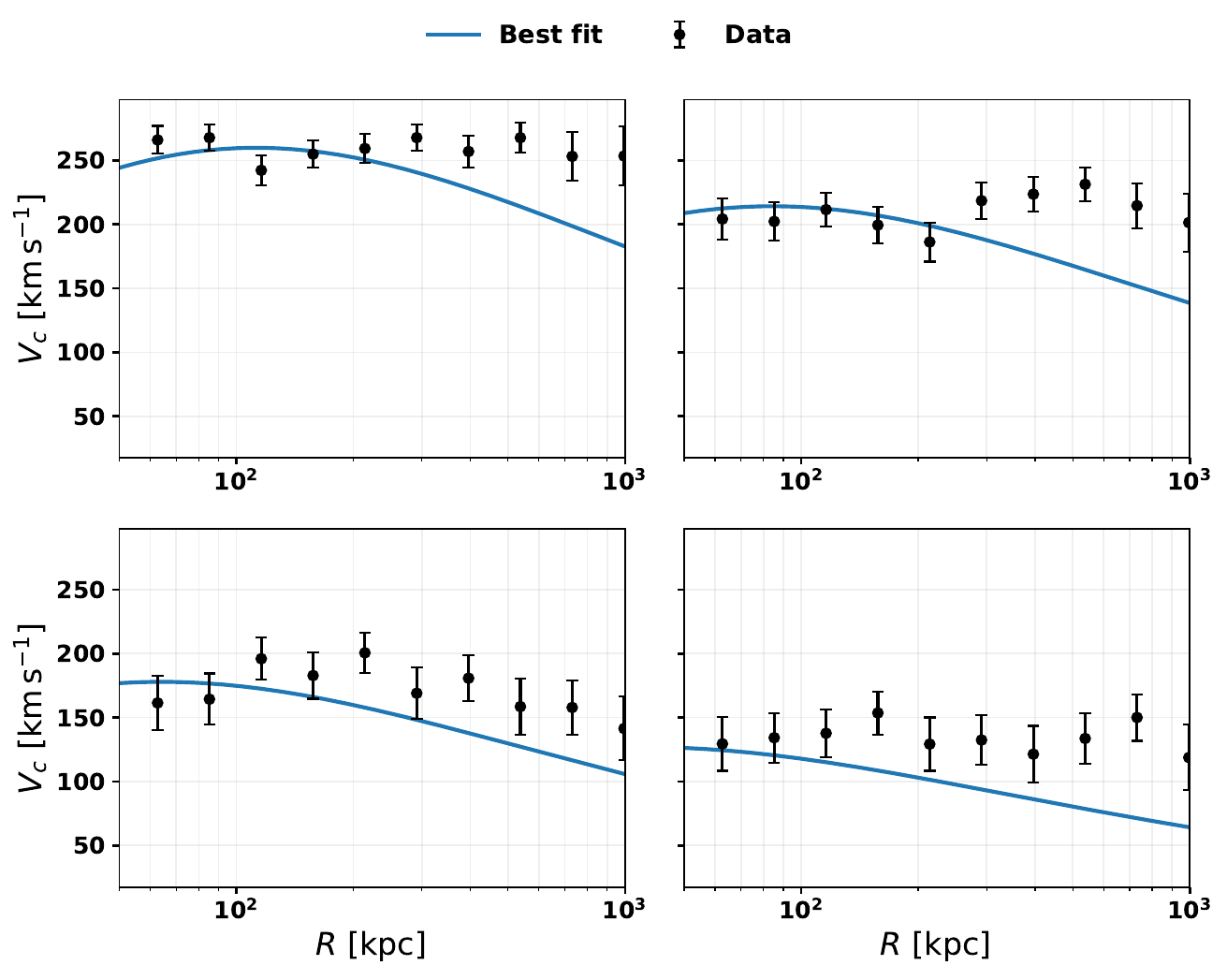}
    \caption{\centering Diemer-Joyce $c$--$M$: $\chi^2_\nu = (4.8,\,5.3,\,2.3,\,3.8)$}
\end{subfigure}

\vspace{0.25cm}

\begin{subfigure}[t]{0.47\textwidth}
    \includegraphics[width=\linewidth]{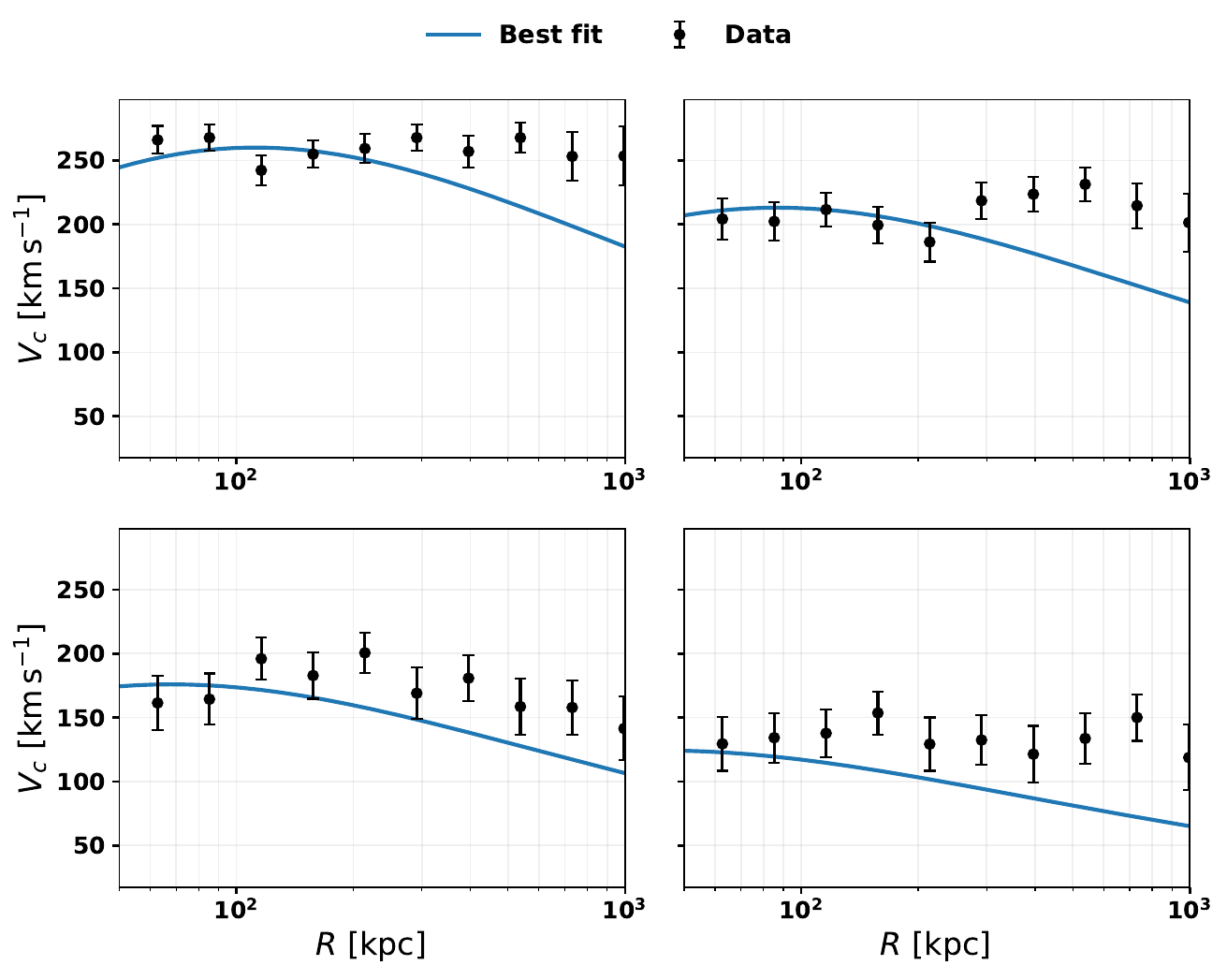}
    \caption{\centering Prada $c$--$M$:
    $\chi^2_\nu = (4.8,\,5.3,\,2.2,\,3.7)$}
\end{subfigure}
\hfill
\begin{subfigure}{0.47\textwidth}
    \includegraphics[width=\linewidth]{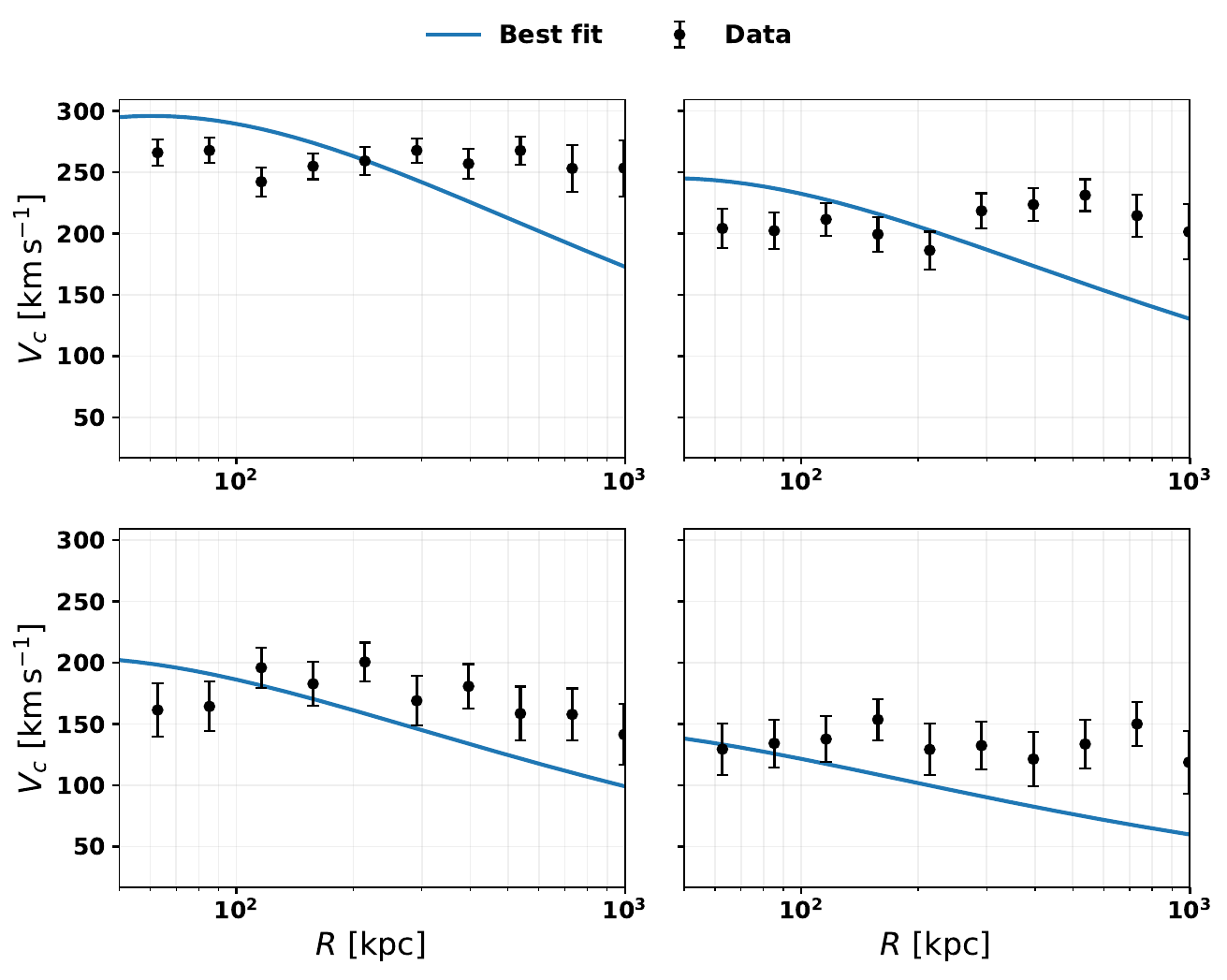}
    \caption{\centering Diemer-Kravtsov $c$--$M$: $\chi^2_\nu = (10.6,\,8.7,\,3.3,\,4.3)$}
\end{subfigure}

\vspace{0.25cm}

\begin{subfigure}{0.47\textwidth}
    \includegraphics[width=\linewidth]{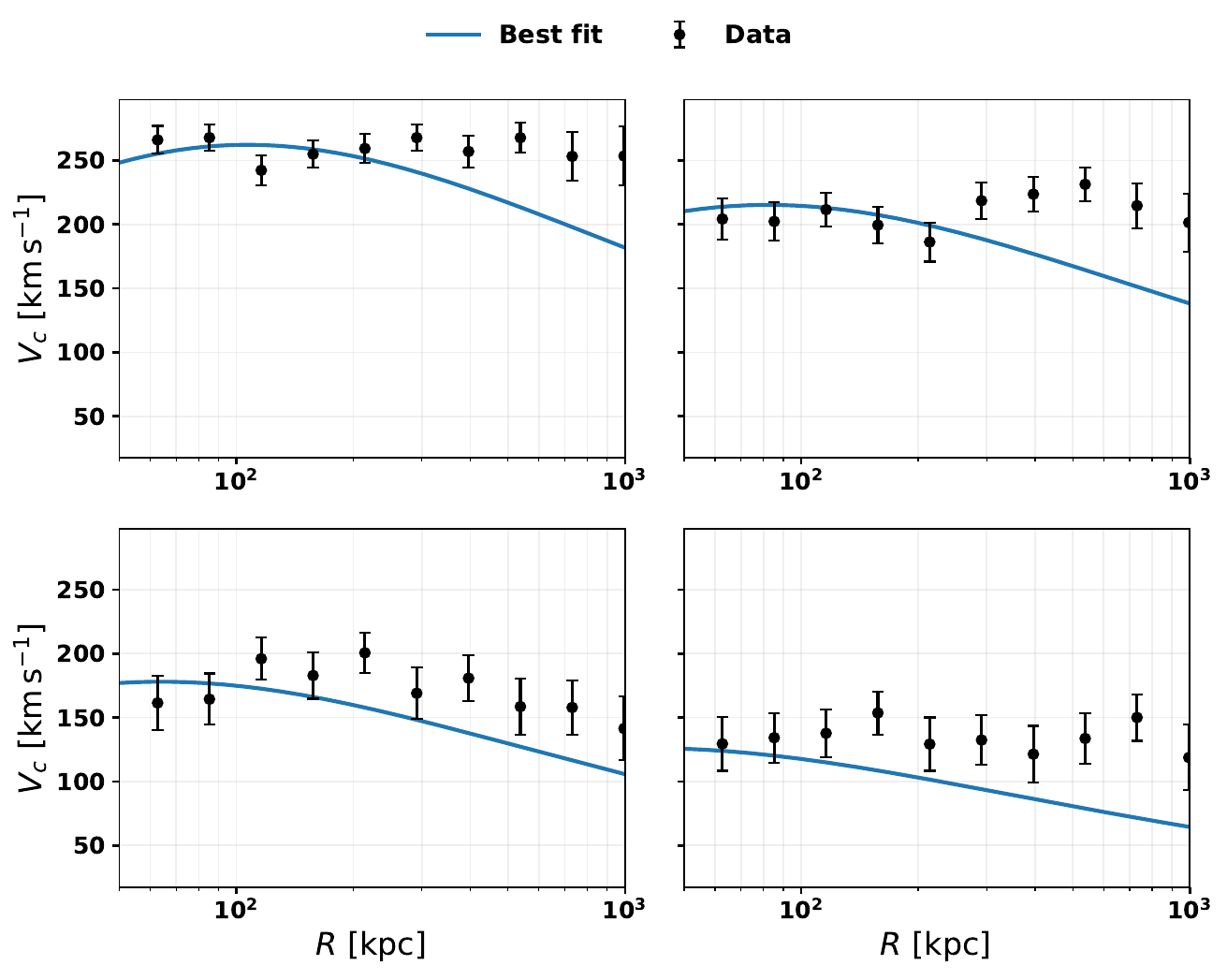}
    \caption{\centering S\'anchez-Conde $c$--$M$: $\chi^2_\nu = (4.9,\,5.4,\,2.3,\,3.7)$}
\end{subfigure}
\hfill
\begin{subfigure}{0.47\textwidth}
    \includegraphics[width=\linewidth]{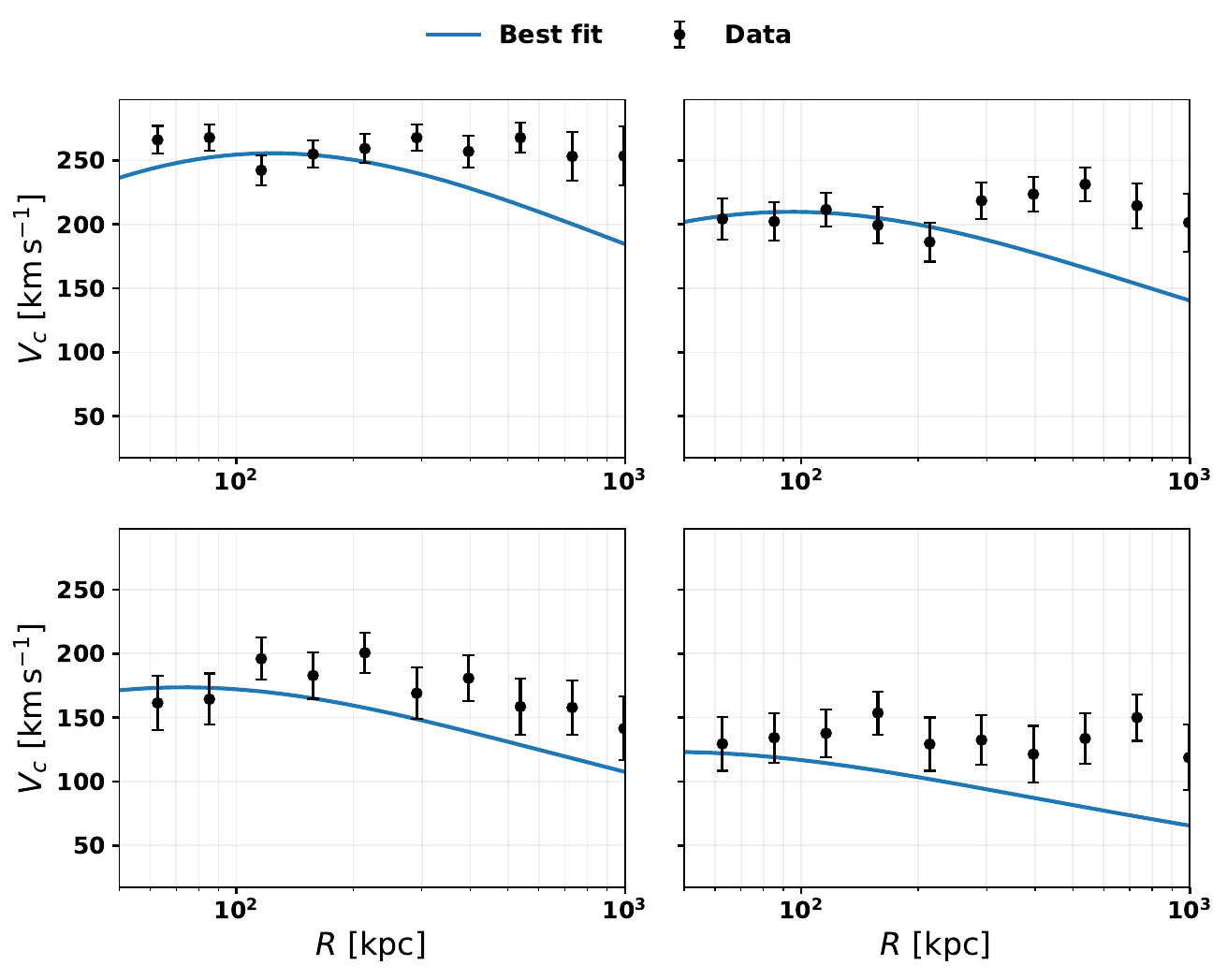}
    \caption{\centering Macci\`{o} $c$--$M$: $\chi^2_\nu = (4.8,\,5.0,\,2.1,\,3.6)$}
\end{subfigure}

\caption{
NFW rotation-curve fits using the Behroozi SMHM relation combined with six different concentration--mass relations.
Within each panel, the four subplots correspond (clockwise from top left) to stellar masses
$M_\star/M_\odot = 1.25\times10^{11}$,
$6.66\times10^{10}$,
$3.46\times10^{10}$,
and $8.47\times10^{9}$.
The reduced chi-squared values for each mass are listed in the individual subfigure captions.
}
\label{fig:behroozi_all}
\end{figure}
\begin{figure}[p]
\captionsetup[subfigure]{justification=centering, singlelinecheck=false, font=small}
\centering

\begin{subfigure}[t]{0.47\textwidth}
    \includegraphics[width=\linewidth]{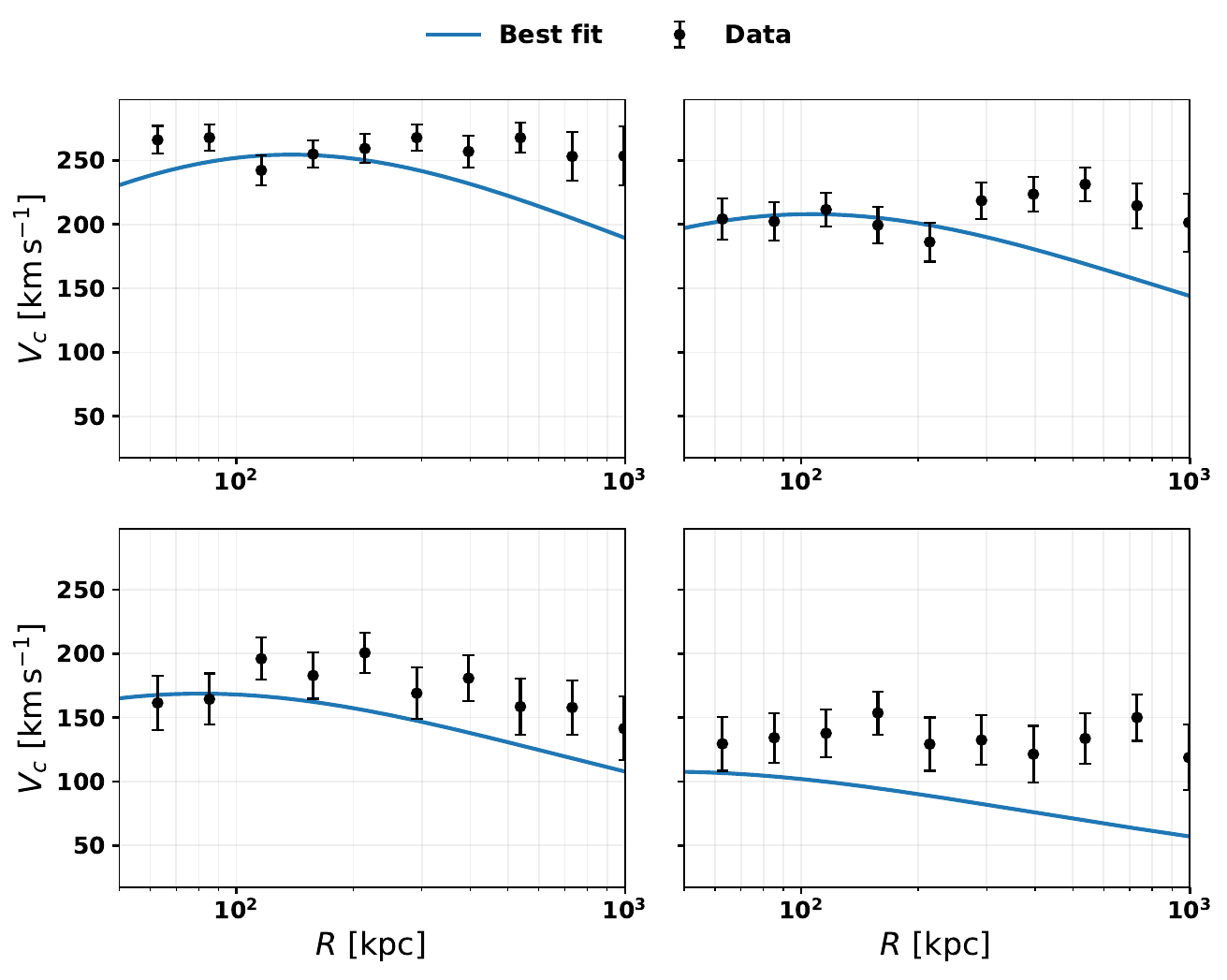}
    \caption{\centering Duffy $c$--$M$: 
    $\chi^2_\nu = (4.4,\,4.5,\,2.2,\,5.7)$}
\end{subfigure}
\hfill
\begin{subfigure}[t]{0.47\textwidth}
    \includegraphics[width=\linewidth]{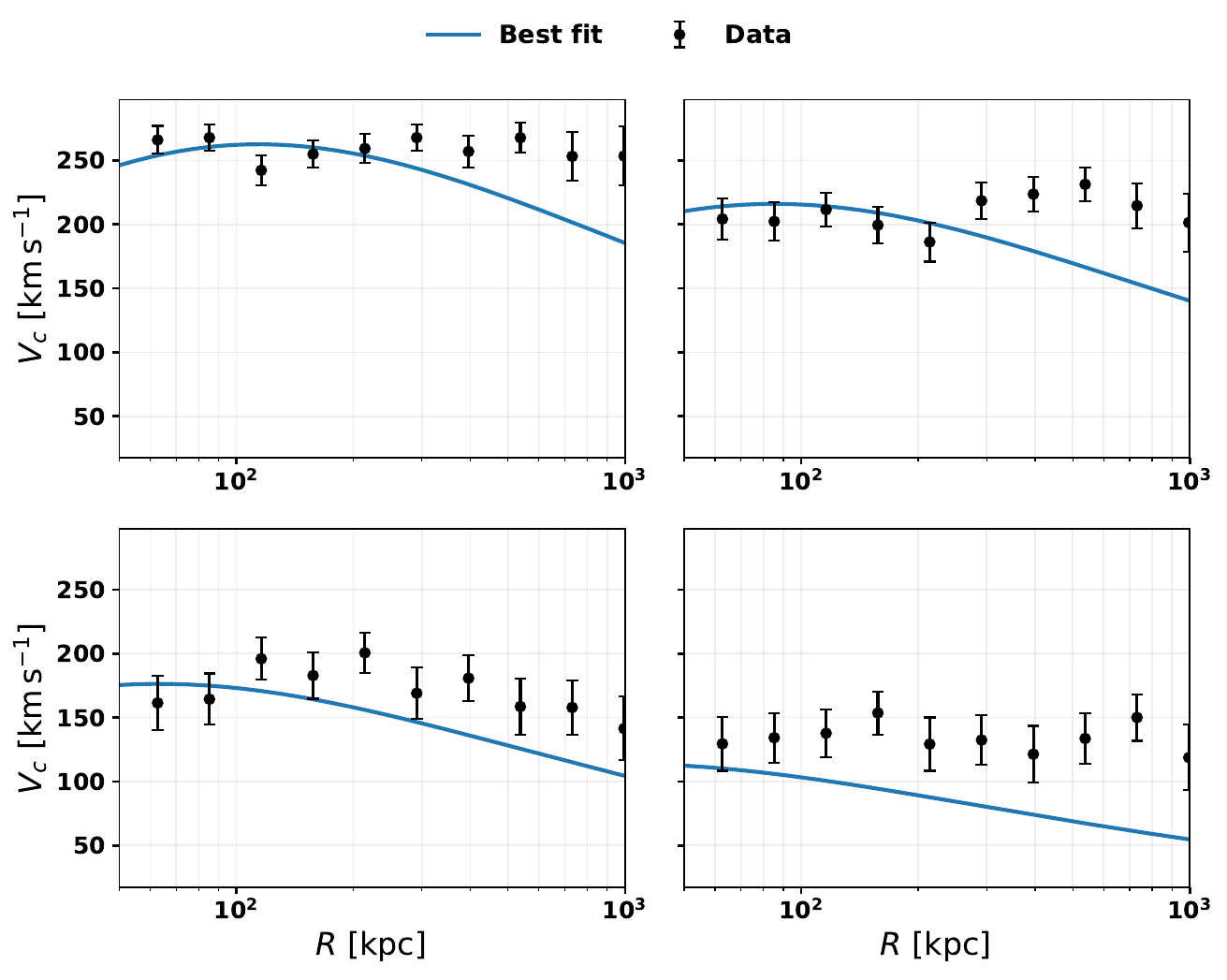}
    \caption{\centering Diemer-Joyce $c$--$M$: $\chi^2_\nu = (4.2,\,5.1,\,2.4,\,5.9)$}
\end{subfigure}

\vspace{0.25cm}

\begin{subfigure}[t]{0.47\textwidth}
    \includegraphics[width=\linewidth]{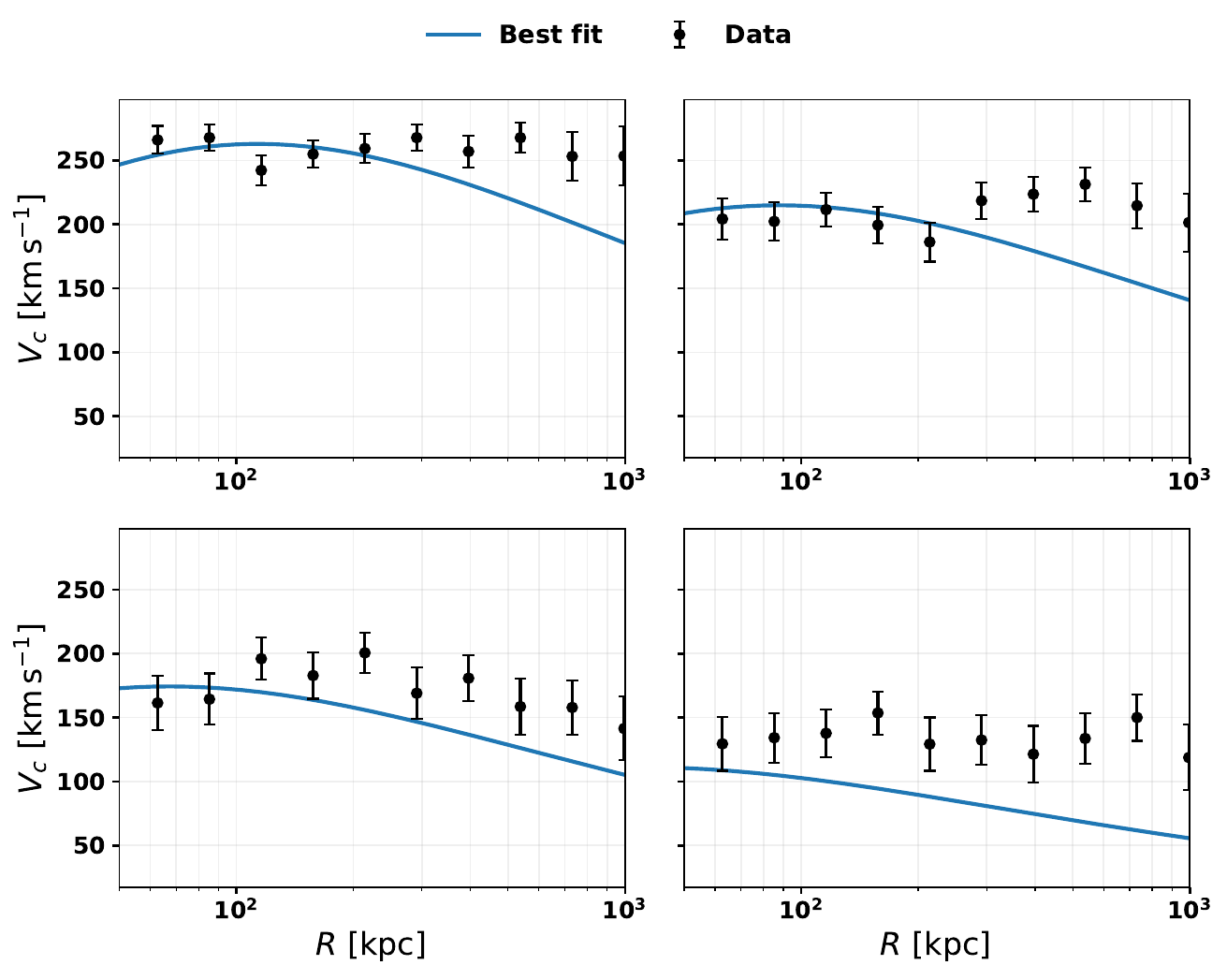}
    \caption{\centering Prada $c$--$M$:
    $\chi^2_\nu = (4.3,\,5.0,\,2.3,\,5.8)$}
\end{subfigure}
\hfill
\begin{subfigure}{0.47\textwidth}
    \includegraphics[width=\linewidth]{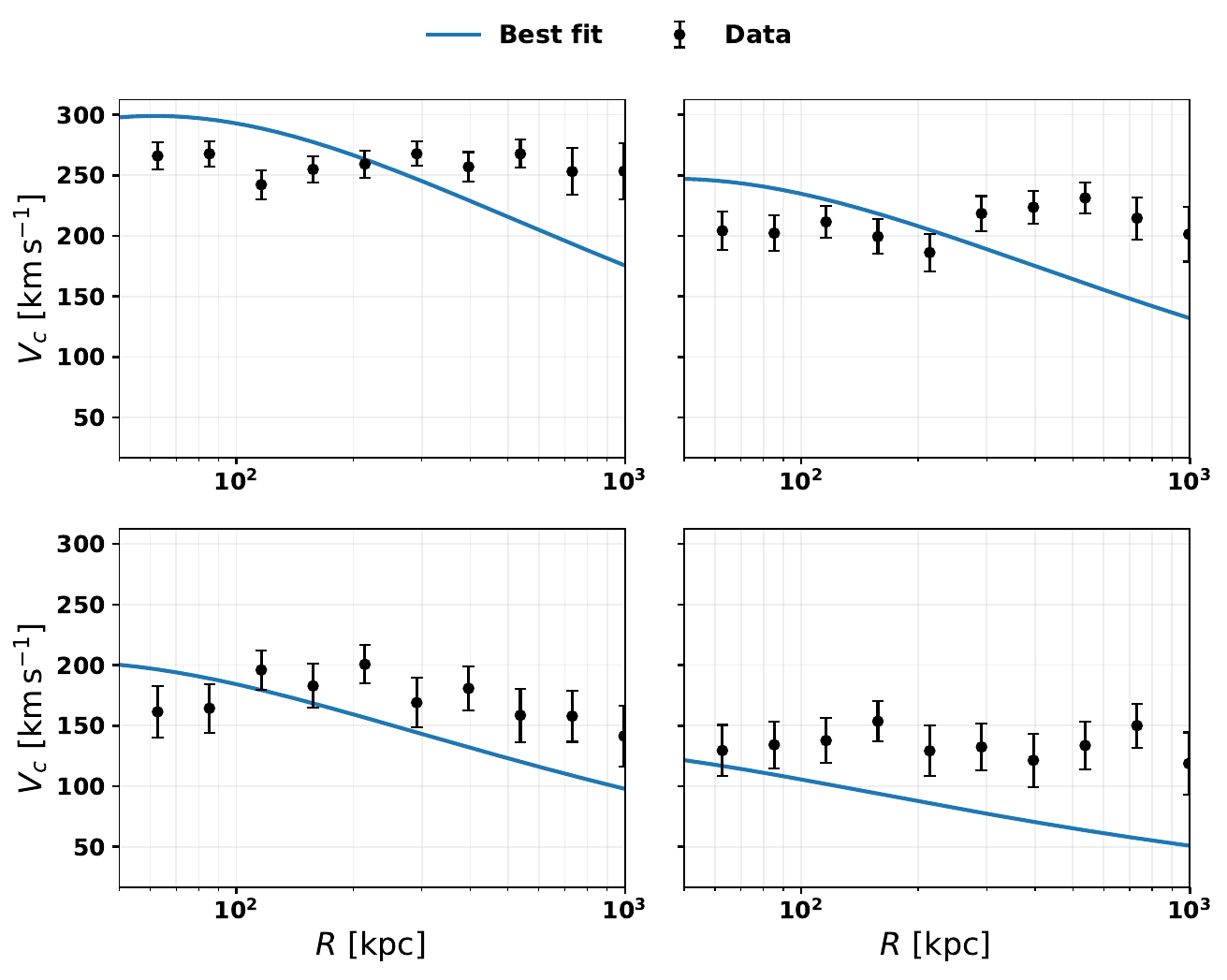}
    \caption{\centering Diemer-Kravtsov $c$--$M$: $\chi^2_\nu = (10.9,\,8.6,\,3.4,\,6.3)$}
\end{subfigure}

\vspace{0.25cm}

\begin{subfigure}{0.47\textwidth}
    \includegraphics[width=\linewidth]{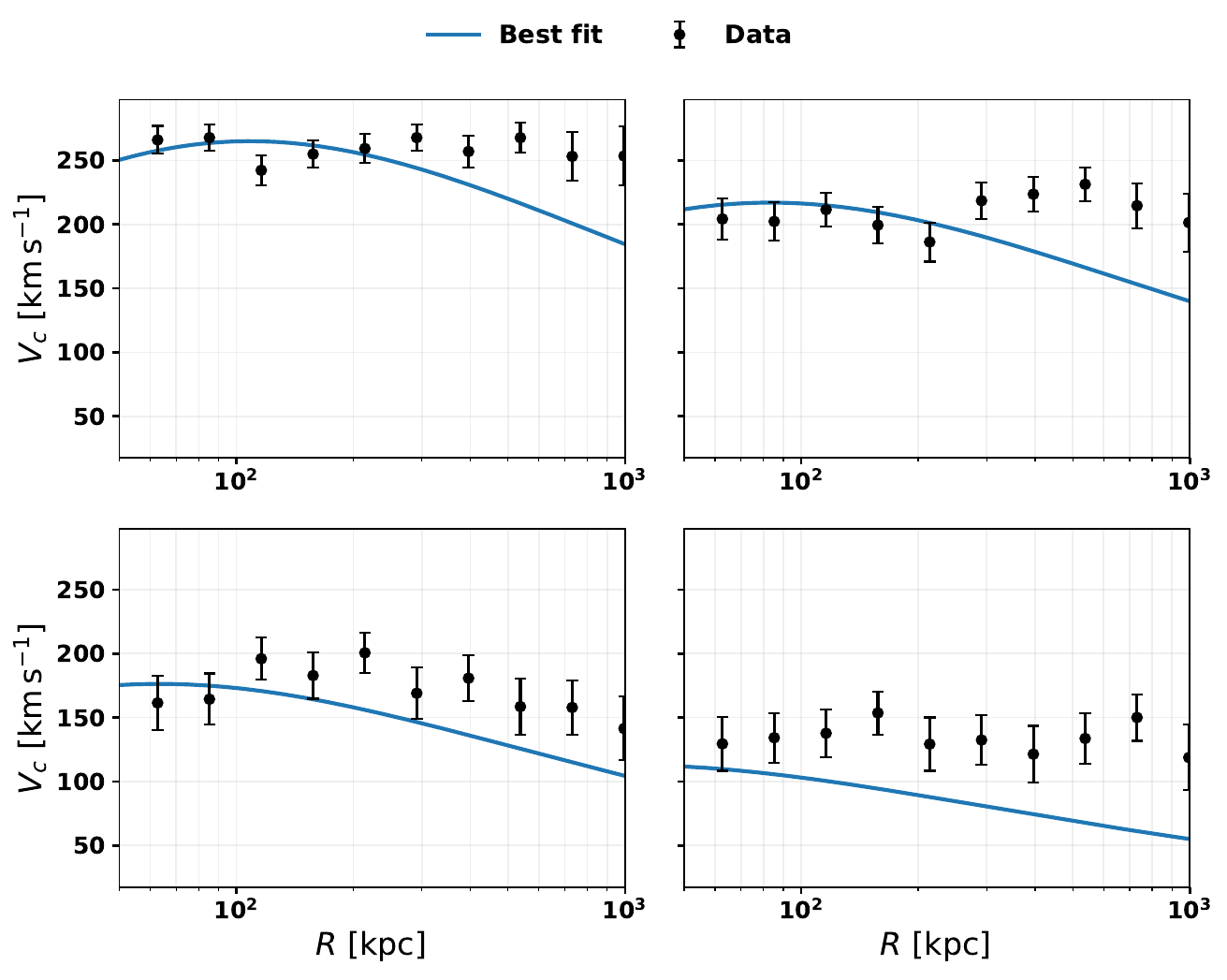}
    \caption{\centering S\'anchez-Conde $c$--$M$: $\chi^2_\nu = (4.4,\,5.2,\,2.4,\,5.9)$}
\end{subfigure}
\hfill
\begin{subfigure}{0.47\textwidth}
    \includegraphics[width=\linewidth]{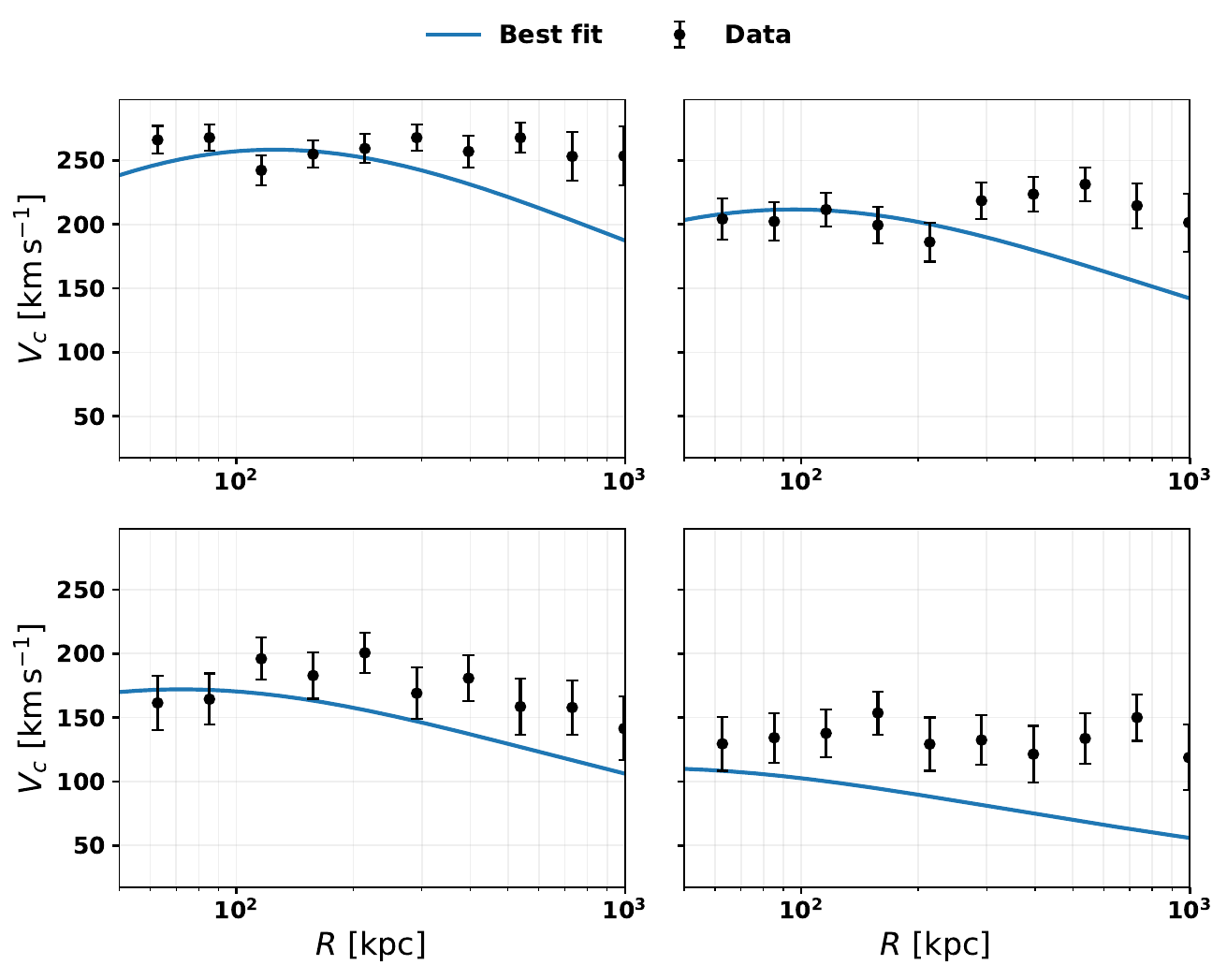}
    \caption{\centering Macci\`{o} $c$--$M$: $\chi^2_\nu = (4.2,\,4.7,\,2.3,\,5.8)$}
\end{subfigure}

\caption{
NFW rotation-curve fits using the Cintio SMHM relation combined with six different concentration--mass relations.
Within each panel, the four subplots correspond (clockwise from top left) to stellar masses
$M_\star/M_\odot = 1.25\times10^{11}$,
$6.66\times10^{10}$,
$3.46\times10^{10}$,
and $8.47\times10^{9}$.
The reduced chi-squared values for each mass are listed in the individual subfigure captions.
}
\label{fig:cintio_all}
\end{figure}

 Representative rotation-curve fits for NFW profile for all $c-M$  and SMHHM  relations are shown   in Figs.~\ref{fig:moster_all}--\ref{fig:cintio_all}. A tabular summary of the reduced $\chi^2$ can be found in Table~\ref{tab:chi2}.
 We find that the reduced $\chi^2$ values are much greater than unity for all combinations of $c-M$ and SHMH relations across all four stellar mass bins. The smallest reduced $\chi^2$ is obtained for the Moster  SHMH and Macci\`{o} $c-M$ relation with values of 2.3, 2.9. 1.6 and 1.3 for the four stellar mass bins. Among the different stellar mass bins, the fourth bin with stellar mass of $8.47 \times 10^9$ $M_{\odot}$ shows minimum $\chi^2$ of less than 2.0 for the Moster SMHM models for all $c-M$ relations except for the Kravtsov relation.
 The maximum discrepancy for most models is observed at large distances at around 1~Mpc. Therefore, the NFW profile cannot adequately describe the lensing profiles for these isolated galaxies.
 However, we should point out that most of the discrepancies between the NFW profile and the data occur at larger radii ($> 300$ kpc). If  the isolation criterion beyond 300 kpc is not robust, the NFW profile can be reconciled  with the observed data by  restricting the fit to radii $< 300$ kpc. This point has also been emphasized in DC25.

\clearpage

\clearpage

\clearpage

\subsection{NFW profile without assuming a $c-M$ relation}
\begin{figure}[h]
\centering
\includegraphics[scale=0.5]{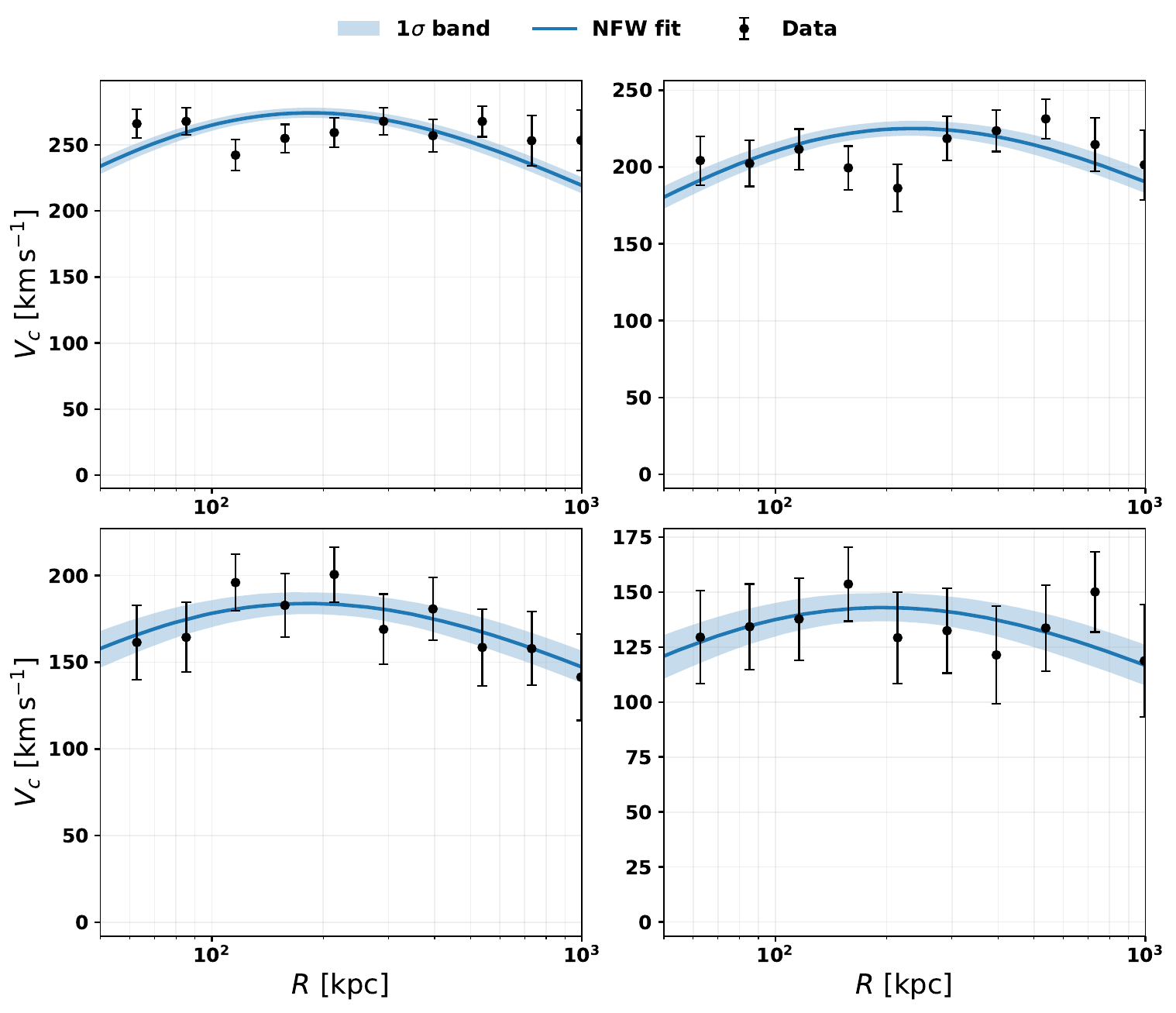}
\caption{\rthis{The best-fit NFW profile without assuming any $c-M$ relation from simulations. Panels shown in clockwise order correspond to stellar masses $M_\star = 1.25 \times 10^{11} M_\odot$, $6.66 \times 10^{10} M_\odot$, $3.46 \times 10^{10} M_\odot$, and $8.47 \times 10^{9} M_\odot$ with $\chi^2_\nu = 2.2$, $1.6$, $0.4$, and $0.4$, respectively. The best-fit parameters can be found in Table~\ref{tab:best}. The shaded regions indicate the theoretical uncertainties in the best-fit parameters.}}
\label{fig:nfwfree}
\end{figure}

\rthis{In addition to assuming a parametric $c-M$ relation, we now do a direct fit to the NFW profile given in Eq.~\ref{eq:nfw} without assuming any $c-M$ or SMHM relation. Therefore similar to Burkert and pseudo-isothermal profiles, $p=2$. The resulting best-fit NFW model together with the data can be found in Fig.~\ref{fig:nfwfree}.  The  reduced $\chi^2$ values are reported in Table~\ref{tab:profile}.
We find that only for the last three stellar mass  bins, the reduced $\chi^2 \lesssim  1.0$. For the first bin, the reduced $\chi^2$ is close to two. Therefore, even without assuming any $c-M$ relation, the NFW profile cannot provide a pristine fit to all the four stellar mass bins. The best-fit values for $\rho_0$ and $r_s$ can be found in Table~\ref{tab:best}. }

\newpage

\subsection{Pseudo-isothermal Profile}
\begin{figure}[h]
\centering
\includegraphics[scale=0.5]{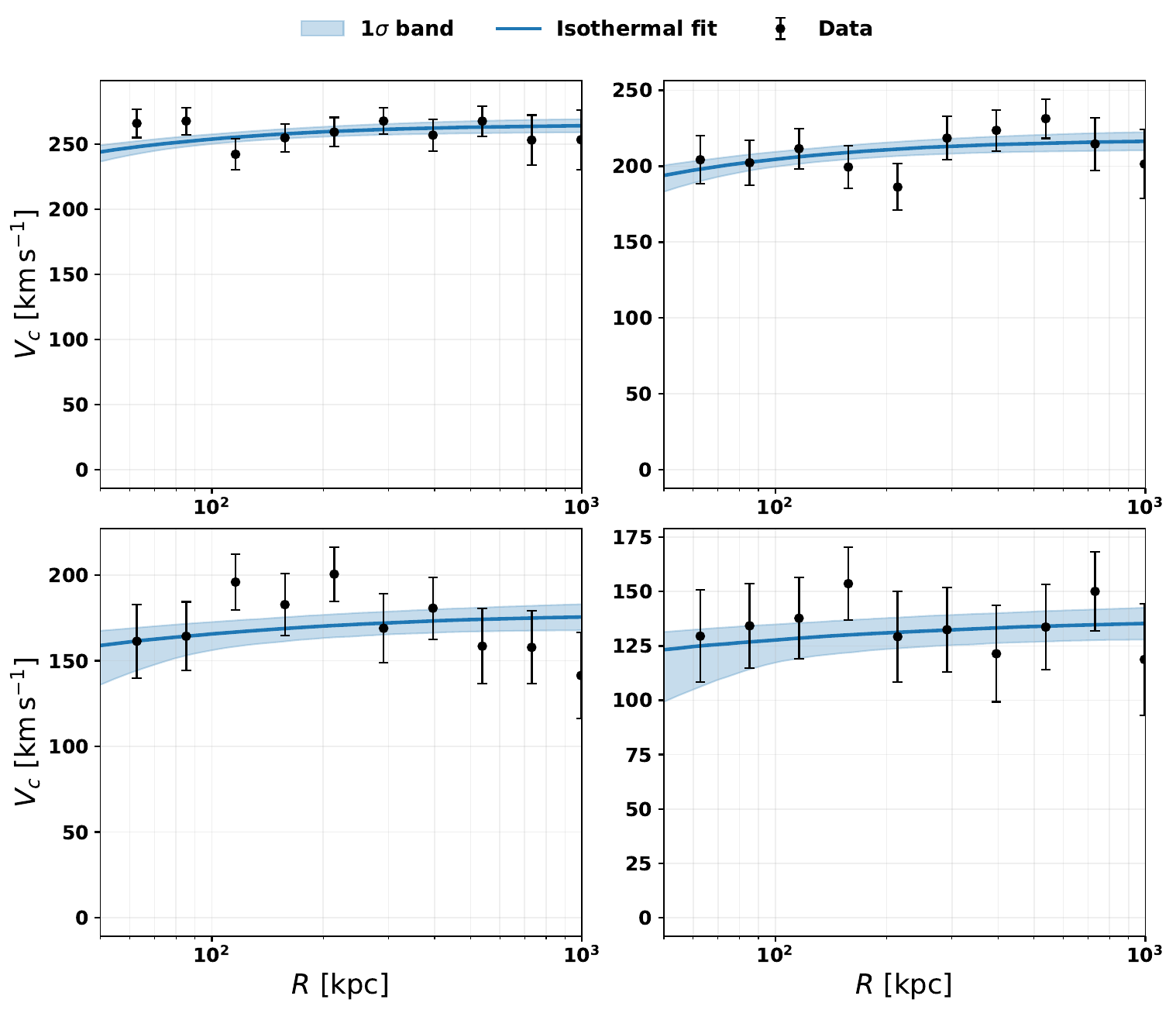}
\caption{The pseudo-isothermal rotation curve \rthis{along with $1\sigma$ uncertainties in the best-fit parameters shown as  shaded regions}. Panels shown in clockwise order correspond to stellar masses $M_\star = 1.25 \times 10^{11} M_\odot$, $6.66 \times 10^{10} M_\odot$, $3.46 \times 10^{10} M_\odot$, and $8.47 \times 10^{9} M_\odot$, where the best-fit  parameters are: $\rho_0 = 5.12 \times 10^7 M_\odot$kpc$^{-3}$ $1.80 \times 10^7M_\odot$kpc$^{-3}$, $1.72 \times 10^7M_\odot$kpc$^{-3}$ and $1.52 \times 10^7M_\odot$kpc$^{-3}$, best-fit $R_c = 5.04$ kpc, $6.98$ kpc, $5.80$ kpc, and $4.76$ kpc, $\chi^2_\nu = 1.0$, $0.8$, $1.3$, and $0.5$ respectively.}
\label{fig:isothermal}
\end{figure}

 The plots showing the best-fit pseudo-isothermal profile  to the circular velocity data  can be found in Fig.~\ref{fig:isothermal} and the reduced $\chi^2$ values have been tabulated in Table~\ref{tab:profile}. We find that the reduced $\chi^2$ is close to  unity for all the four stellar mass bins. \rthis{The best-fit values for the free parameters can be found in Table~\ref{tab:best}.}

\newpage

\subsection{Burkert Profile}
\begin{figure}[h]
\centering
\includegraphics[scale=0.5]{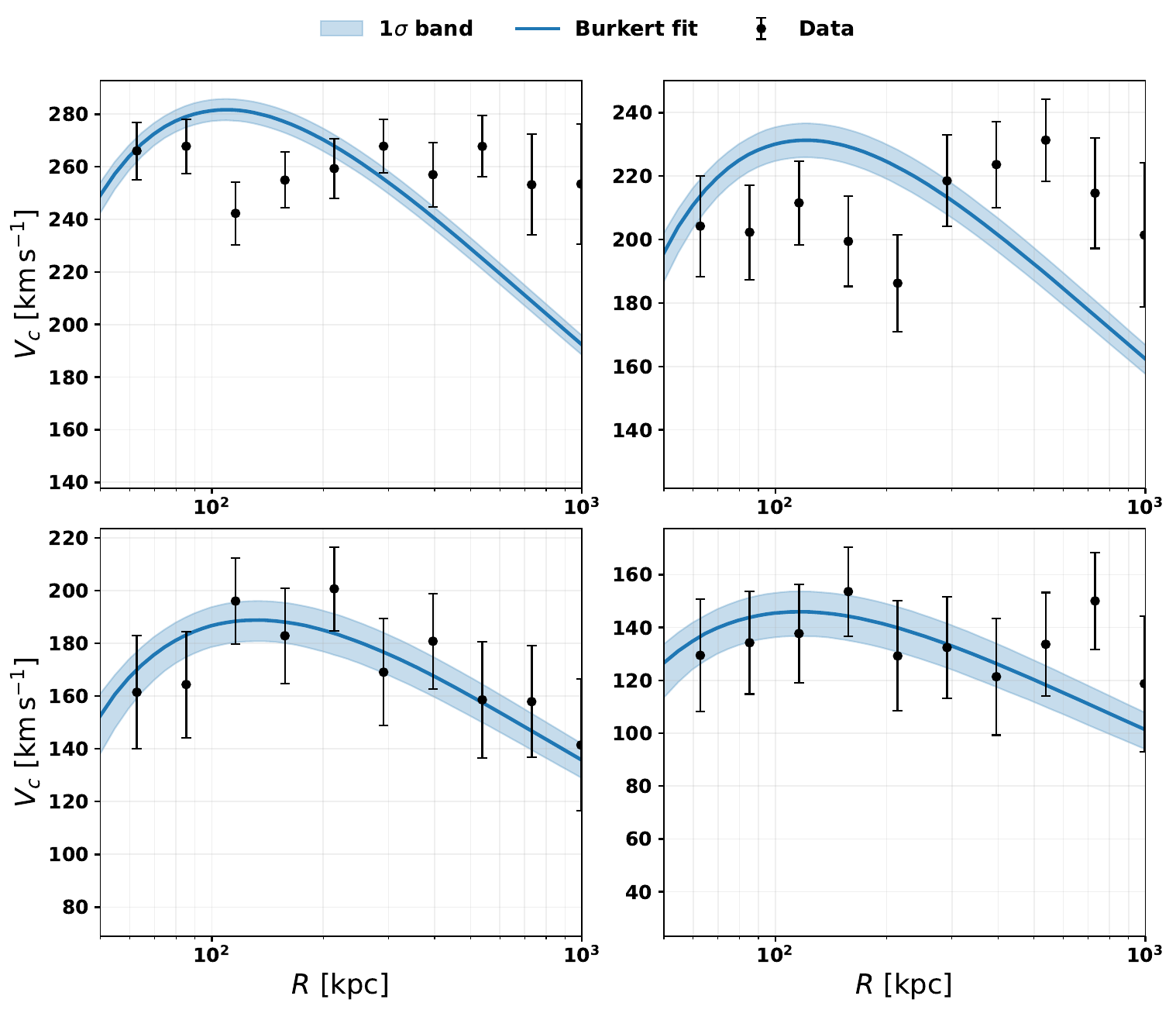}
\caption{The Burkert rotation curve \rthis{along with $1\sigma$ uncertainties in the best-fit parameters shown as  shaded regions}. Panels shown in clockwise order correspond to stellar masses $M_\star = 1.25 \times 10^{11} M_\odot$, $6.66 \times 10^{10} M_\odot$, $3.46 \times 10^{10} M_\odot$, and $8.47 \times 10^{9} M_\odot$, 
with best-fit parameters: $\rho_0 = 3.03 \times 10^7 M_\odot$kpc$^{-3}$ $1.67 \times 10^7M_\odot$kpc$^{-3}$, $9.43 \times 10^6M_\odot$kpc$^{-3}$ and $7.36 \times 10^6M_\odot$kpc$^{-3}$, best fit $r_c = 15.94$ kpc, $17.66$ kpc, $19.39$ kpc, and $17.13$ kpc, $\chi^2_\nu = 5.6$, $3.8$, $0.3$, and $0.7$ respectively.}
\label{fig:burkert}
\end{figure}

The  best-fit Burkert profiles along with the data are shown in Fig.~\ref{fig:burkert}. The reduced $\chi^2$ values are tabulated in Table~\ref{tab:profile}. \rthis{The best fit values for the free parameters  can be found  in Table~\ref{tab:best}.}

 We find that the first two stellar mass bins ($1.25 \times 10^{11} M_{\odot}$ and $6.66 \times 10^{10} M_{\odot}$) are poorly fitted with reduced $\chi^2>3$. However, for the last two stellar mass bins, the reduced $\chi^2$ values are less than 1.0. Thus, the Burkert profile  does not provide a robust fit across all four stellar mass bins.

 Therefore, among the three dark matter profiles, only the pseudo-isothermal profile provides a robust fit to the circular velocity profiles for all the four  stellar mass bins.


\begin{table*}[h]
\centering
\caption{Reduced $\chi^2$ values for NFW profile across all  $c-M$ and SMHM relations for the four  stellar mass bins.}
\label{tab:chi2}
\begin{tabular}{|c|c|c|c|c|c|}
\hline
\textbf{SMHM Model} &
\textbf{$c$--$M$ Model} &
\boldmath$\chi^2_\nu$ &
\boldmath$\chi^2_\nu$ &
\boldmath$\chi^2_\nu$ &
\boldmath$\chi^2_\nu$ \\
& & \boldmath$1.25\times10^{11}\,M_\odot$ & \boldmath$6.66\times10^{10}\,M_\odot$ & \boldmath$3.46\times10^{10}\,M_\odot$ & \boldmath$8.47\times10^{9}\,M_\odot$ \\
\hline
Moster \cite{Moster13} & Duffy  \cite{2008MNRAS.390L..64D}  & 2.7  & 3.1 & 1.8 & 1.4 \\
 & Diemer-Joyce \cite{2019ApJ...871..168D}  & 3.2  & 5.1 & 2.0 & 1.6 \\
 & Prada \cite{2012MNRAS.423.3018P}  & 4.2  & 4.1 & 1.9 & 1.5 \\
 & Diemer-Kravtsov \cite{2015ApJ...799..108D} & 3.2  & 7.0 & 3.4 & 2.5 \\
 & S\'{a}nchez-Conde \cite{2014MNRAS.442.2271S} & 4.8  & 4.4 & 2.0 & 1.9 \\
 & Macci\`{o} \cite{Maccio08} & 2.3 & 2.9 & 1.6 & 1.3 \\
\hline
Behroozi \cite{2013ApJ...770...57B} & Duffy   \cite{2008MNRAS.390L..64D}  & 5.1  & 4.8 & 2.0 & 3.6 \\
 & Diemer-Joyce \cite{2019ApJ...871..168D} & 4.8  & 5.3 & 2.3 & 3.8 \\
 & Prada  \cite{2012MNRAS.423.3018P}  & 4.8  & 5.3 & 2.2 & 3.7 \\
 & Diemer-Kravtsov \cite{2015ApJ...799..108D} & 10.6 & 8.7 & 3.3 & 4.3 \\
 & S\'{a}nchez-Conde\cite{2014MNRAS.442.2271S} & 4.9  & 5.4 & 2.3 & 3.7 \\
 & Macci\`{o} \cite{Maccio08} & 4.8 & 5.0 & 2.1 & 3.6 \\
\hline
Cintio \cite{2014MNRAS.441.2986D} & Duffy  \cite{2008MNRAS.390L..64D}   & 4.4  & 4.5 & 2.2 & 5.7 \\
 & Diemer-Joyce \cite{2019ApJ...871..168D}  & 4.2  & 5.1 & 2.4 & 5.9 \\
 & Prada  \cite{2012MNRAS.423.3018P}  & 4.3  & 5.0 & 2.3 & 5.8 \\
 & Diemer-Kravtsov \cite{2015ApJ...799..108D} & 10.9 & 8.6 & 3.4 & 6.3 \\
 & S\'{a}nchez-Conde \cite{2014MNRAS.442.2271S} & 4.4  & 5.2 & 2.4 & 5.9 \\
 & Macci\`{o} \cite{Maccio08} & 4.2 & 4.7 & 2.3 & 5.8 \\
\hline
\end{tabular}
\end{table*}


\begin{table}[h]
\centering
\caption{Reduced $\chi^2$ values for the NFW \rthis{(without assuming any $c-M$ relation)}, pseudo-isothermal and Burkert profiles.}
\label{tab:profile}
\renewcommand{\arraystretch}{1.4}
\begin{tabular}{|c|c|c|c|c|}
\hline
\textbf{Dark Matter Profile} &
\boldmath$\chi^2_\nu$ &
\boldmath$\chi^2_\nu$ &
\boldmath$\chi^2_\nu$ &
\boldmath$\chi^2_\nu$ \\
    & \boldmath$1.25\times10^{11} M_\odot$ & \boldmath$6.66\times10^{10} M_\odot$ & \boldmath$3.46\times10^{10} M_\odot$ & \boldmath$8.47\times10^{9} M_\odot$ \\
\hline
NFW  (without $c-M$ relation)  & 2.2 & 1.6 & 0.4 & 0.4 \\
isothermal \cite{Moster13} & 1.0 & 0.8 & 1.3 & 0.5 \\
Burkert \cite{Moster13} & 5.6 & 3.8 & 0.3 & 0.7 \\
\hline
\end{tabular}
\end{table}

\begin{table}[h]
\centering
\caption{Best fit values for the \rthis{NFW (without assuming any $c-M$ relation)}, pseudo-isothermal and Burkert profiles.}
\label{tab:best}
\renewcommand{\arraystretch}{1.4}
\begin{tabular}{|c|c|c|c|c|c|}
\hline
\textbf{Profile} & \textbf{Best fit Parameter} &
    \boldmath$1.25\times10^{11} M_\odot$ & \boldmath$6.66\times10^{10} M_\odot$ & \boldmath$3.46\times10^{10} M_\odot$ & \boldmath$8.47\times10^{9} M_\odot$ \\
\hline
NFW  & $\rho_0 (\times 10^5 M_\odot\text{kpc}^{-3})$ & $8.94\pm1.6 $ & $3.72\pm1.0$ & $4.13\pm2.0$ & $2.18\pm1.2$ \\
(without $c-M$ relation) & $r_s (\text{kpc})$ & $84.92\pm7.93$ & $108.07\pm14.74$ & $84.14\pm19.62$ & $90.14\pm25.39$ \\
\hline
pseudo-isothermal & $\rho_0 (\times 10^7 M_\odot\text{kpc}^{-3})$ & $5.12\pm4.4 $ & $1.80\pm1.8$ & $1.72\pm3.8$ & $1.52\pm4.6$ \\
 & $R_c (\text{kpc})$ & $5.04\pm2.24$ & $6.98\pm3.63$ & $5.80\pm6.63$ & $4.76\pm7.37$ \\
\hline
Burkert & $\rho_0 (\times 10^6 M_\odot\text{kpc}^{-3})$ & $30.3\pm3.0$ & $16.74\pm2.3$ & $9.43\pm2.1$ & $7.36\pm2.1$ \\
 & $r_c (\text{kpc})$ & $15.94\pm0.76$ & $17.66\pm1.16$ & $19.392.05$ & $17.13\pm2.39$ \\
\hline
\end{tabular}
\end{table}


\section{Conclusions}
\label{sec:conclusions}
In this work, we have revisited the compatibility of three dark matter profiles  with stacked weak lensing based circular velocity measurements from the KiDS survey, which have previously been discussed in two recent works (M24 and DC25). M24  had found that the rotation curves were flat up to 1 Mpc, which at first glance is not consistent with the NFW profile since  it predicts a falling rotation curve. DC25 argued that if one uses $M_{100}$ instead of $M_{200}$ as a proxy for the virial mass in conjunction with  a different SMHM relation, one can get better agreement with the NFW profile.

We extended the analysis in M24 by considering six different $c-M$ relations and four SMHM relations while fitting the NFW profile to the circular velocity data. \rthis{We also  fitted  the NFW profile without assuming any $c-M$ relation.}
In addition we also fitted the cored Burkert and pseudo-isothermal profile. We find that the reduced $\chi^2$ values are greater than 1.0 for almost all combinations of SMHM and $c-M$ relations.  The Burkert  profile provides poor fits to  the first two stellar mass bins but can adequately fit the last two stellar mass bins. Only the pseudo-isothermal profile provides a robust fit to all the four stellar mass bins. 

\begin{acknowledgments}
We are grateful to Man-Ho Chan for providing the circular  velocity data from DC25 and M24 used for this analysis.
\end{acknowledgments}
\bibliography{apssamp}
\end{document}